\documentclass[iop,english]{emulateapj}
\usepackage{float}
\usepackage{amssymb}
\usepackage{graphicx}
\usepackage{babel}
\begin{document}

\title{Secular dynamics in hierarchical three-body systems\\
with mass loss and mass transfer}

\author{Erez Michaely and Hagai B. Perets}

\affil{Physics Department, Technion - Israel Institute of Technology, Haifa
3200004, Israel}
\begin{abstract}
Recent studies have shown that secular evolution of triple systems
can play a major role in the evolution and interaction of their inner
binaries. Very few studies explored the stellar evolution of triple
systems, and in particular the mass loss phase of the evolving stellar
components. Here we study the dynamical secular evolution of hierarchical
triple systems undergoing mass loss. We use the secular evolution
equations and include the effects of mass-loss and mass-transfer,
as well as general relativistic effects. We present various evolutionary
channels taking place in such evolving triples, and discuss both the
effects of mass-loss and mass-transfer in the inner binary system,
as well as the effects of mass-loss/transfer from an outer third companion.
We discuss several distinct types/regimes of triple secular evolution,
where the specific behavior of a triple system can sensitively depend
on its hierarchy and the relative importance of classical and general
relativistic effects. We show that the orbital changes due to mass-loss
and/or mass-transfer processes can effectively transfer a triple system
from one dynamical regime to another. In particular, mass loss/transfer
can both induce and quench high amplitude (Lidov-Kozai) variations
in the eccentricity and inclination of the inner binaries of evolving
triples. They can also change the system dynamics from an orderly
periodic behavior to a chaotic one, and vice versa. 
\end{abstract}

\section{Introduction}

Triple systems are some of the most frequent astrophysical phenomena,
manifesting themselves in almost any given scale; where triple stars,
planets in binaries, and even our own Sun-Earth-moon system serve
as a few obvious examples. About 15 \% of all stars reside in triples
(\citealt{Raghavan2010}; and possibly$>$50 (40) \% for more massive
O/B stars; \citealt{Tokovinin1997,Eggleton2007,RemageEvans2011};
S. De-Mink, private communication, 2014). Some aspects of the rich
and complex gravitational dynamics of such systems have been studied
extensively (as evidenced by the well known three body problem; \citealt{Valtonen2006}).
However, study of the realistic evolution of such systems, including
the coupling of their dynamics with non-gravitational processes and/or
the realistic treatment of the physical properties such as mass loss
and mass transfer is still in its infancy. Given the mounting evidence
for the importance of such systems, and their ubiquity in stellar
and planetary systems, our poor understanding of these systems is
quite disconcerting. Here we explore the secular dynamics of evolving
triples, and in particular the effects of mass-loss and mass-transfer
on the dynamical evolution of triples. 

Stable triple systems are hierarchical, namely, consists of an inner
binary and an outer binary orbit, i.e. the tertiary (third object).
Secular evolution of such systems is the change of \textit{orbital
elements} on timescale much larger than the dynamical timescale of
the system. Triple stellar evolution processes, i.e. mass loss, mass
transfer, tidal friction, gravitational wave emission, etc. influence
the orbital elements through changes in energy and angular momentum.
Hence, in order to understand triple stellar evolution one should
couple the secular dynamics and the evolutionary processes to get
a complete and more comprehensive picture. 

Several astrophysical phenomena are likely to be produce via triple
interaction and evolution. Key dynamical long term effect is the Lidov-Kozai
mechanism \citet{Kozai1962,Lidov1962}, in which perturbations by
the third outer companion lead to periodic/quasi-periodic (and sometime
chaotic; \citet{Ford2000,Blaes2002,Katz2012,Naoz2013a} ) large amplitude
oscillations (sometime called Kozai-cycles) of the mutual inclination
of the inner and outer binary in the triple, as well as the inner
binary eccentricity. This secular effect, first introduced in the
context of Solar system bodies, have been since suggested to play
an important role over a wide range of systems and scales, from the
dynamics of super-massive black holes and stars to exoplanets, moons
and planetesimals \citep[e.g.][]{Kozai1962,Lidov1962,Blaes2002,Nesvorny2003,Eggleton2006,Wu2003,Fabrycky2007,Perets2009b,Thompson2011,Antonini2012,Katz2012,Naoz2012,Naoz2013}.
The basic effect of the Kozai cycle is that on time scale much larger
than the orbits periods, the inner orbit eccentricity and relative
inclination fluctuate due to mutual torque between the inner and outer
binaries. As a result, orbits exchange angular momentum (but not energy)
and become more eccentric, possibly to the point in which physical
collisions/mergers, tidal friction, gravitational wave emission and/or
other short range effects become important. In the following we consider
all the system components to be point like mass particles; the coupling
of tidal effects and/or mergers/collisions where the physical size
plays a role will be explored elsewhere.

The analysis of secular processes make use of the mass averaging technique;
the masses of the triple components are averaged-out over the inner
and outer binaries orbital periods, and the dynamics then follow the
mutual torques of these mass-averaged rings, using a perturbative
expansion. For a very low-mass secondary in the inner binary (the
test-particle regime) using expansion terms up to the quadruple level
is sufficient to describe its evolution (e.g. for the cases of a star-planet-comet
or a star-planet-moon systems; \citealt{Naoz2013}). The quadruple
expansion is also sufficient in symmetric cases, where the primary
and secondary in the inner binary have equal masses and the octupole
expansion term becomes negligible (see later section for quantitative
description). Octupole expansion treatment exhibits features lacking
in the quadrupole analysis. In particular, system with no inner binary
symmetry in mass i.e. $m_{0}\neq m_{1}$ where $m_{0}$ and $m_{1}$
are the masses of the inner binary \citep{Mazeh1979,Ford2000,Blaes2002,Naoz2013}.
Octupole level dynamical treatment covers a wider range of systems,
but it becomes inaccurate once the triple hierarchy is weaker, i.e.
the timescales for the third body perturbation during periastron passage
becomes comparable to the dynamical timescale of the inner binary,
and the secular averaging approach is no longer valid \citep{Antonini2012,Katz2012,Antognini2014,Antonini2014,Prodan2014}.
Each one of these approaches is also characterized by a different
dynamical behavior, from a periodical behavior in the quadrupole regime,
to a quasi-periodic and chaotic behavior at the octupole and the non-secular
regime.

In the following we use the secular approach to explore mass transfer
and mass loss processes in triple systems. This approach is valid
as long as the mass-loss/transfer processes occur on long enough timescales,
such that the changes in angular momentum/energy on dynamical timescales
are small; prompt mass-loss processes such as supernovae explosions
are therefore not studied here. This is done by adding the appropriate
terms for mass loss and mass transfer to the secular equations of
motion at the octupole level, as well as using direct N-body integration
to validate our secular analysis. As we show, such processes can change
the dynamical regime characterizing the system, and thereby lead to
significant changes in the system behavior. We focus only on mass-loss
induced transitions in the secular regimes (see also an N-body study
of such transitions by \citealt{Shappee2013}); discussion of mass-loss
induced transition from the secular regime to non-secular or even
unstable triple regime is discussed elsewhere \citep[e.g.][]{Kratter2012,Perets2012,Veras2012}.
For related studies of mass-loss in multi-planet (orbiting single
stars) see \citet{Voyatzis2013}.

Our paper is organized as follows: In Sec. \ref{sec:Octupole-order-of}
we briefly review the standard octupole order expansion and describe
the equations of motion; readers familiar with this approach may skip
to Sec. \ref{sec:Secular-evolution-with-massloss} where we present
the additional secular mass loss/transfer terms to the equation of
motion. In Sec. \ref{sec:Evolutionary-Channels} we provide detailed
examples for the evolution of triple systems, and compare the results
with direct N-body simulations. We then (Sec. \ref{sec:Discussion-and-Summary})
discuss novel triple evolutionary channels accessible due to the coupling
of secular dynamics with mass loss/mass transfer.

\section{Octupole order of secular evolution}

\label{sec:Octupole-order-of}In the following we first briefly present
the analytic treatment for the long term, secular evolution of the
hierarchical triple systems following previous studies \citep{Harrington1968,Ford2000,Naoz2013,Mazeh1979}.
These provide the basis for the secular evolution approach, which
we then extended to include mass-loss and mass-transfer processes.
We use time-independent Hamiltonian perturbation theory and expand
it up to the third, octupole order. 

\begin{figure}[h]
\includegraphics[scale=0.4]{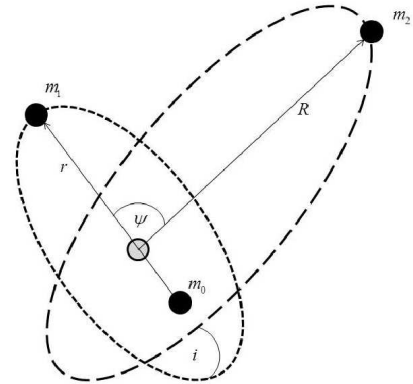}

\caption{\label{fig:Heirarchichal-theree-body}Hierarchical three body system.
Two weakly interacting binaries: inner Keplarian orbit (binary) consists
of mass $m_{0}$ and $m_{1}$ and an outer Keplarian orbit (binary)
of $m_{2}$ and the center of mass of the inner binary, denoted in
a Grey circle. $r$ - is the separation vector of the inner binary;
$R$- is the separation vector of the outer binary; $i$ - is the
inclination between the two orbiting planes and $\psi$ is the angle
berween the separation vectors $r$ and $R$. }
\end{figure}

One can treat an hierarchical triple as two weakly interacting Kepler
orbits, an inner orbit and an outer orbits (see Fig. \ref{fig:Heirarchichal-theree-body})
together with a weak interaction term. We than exploit the fact that
an hierarchical system, by definition, satisfies the following condition
\begin{equation}
\alpha\equiv\frac{a_{1}}{a_{2}}\ll1\label{eq:alpha_definition}
\end{equation}
where $a_{1}$ is the inner binary semi major axis (SMA) and $a_{2}$
is the outer binary SMA. The interaction term can then be expanded
by powers of $\alpha$, as we show in the following. 

\begin{figure}[H]
\includegraphics[scale=0.5]{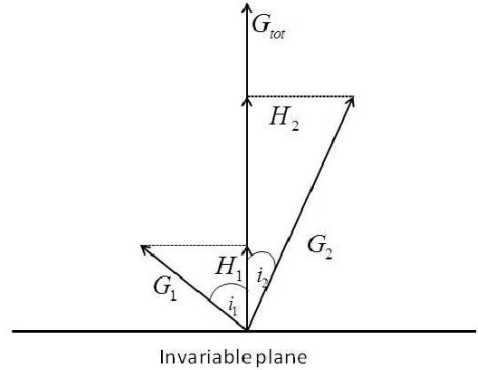}

\caption{\label{fig:Delaunays elements}Geometry of the angular momenta of
the system. $G_{1}$ is the orbital angular momentum of the inner
binary. $G_{2}$ is the orbital angular momentum of the outer binary.
$H_{1}$ is the z-component of the angular momentum of the inner binary.
$H_{2}$ is the z-component of the angular momentum of the outer binary.
The total angular momentum is $H=G_{tot}=G_{1}+G_{2}$. The angle
between $G_{1}$ and $G_{tot}$ is $i_{1}$ and respectively $i_{2}.$
The mutual inclination is $i=i_{1}+i_{2}.$}
\end{figure}

The Hamiltonian of a triple system (not necessarily an hierarchical
one) is given by 
\begin{equation}
\mathcal{H}=\frac{Gm_{0}m_{1}}{2r}+\frac{G\left(m_{0}+m_{1}\right)m_{2}}{2R}+V\left(r,r_{12},r_{23}\right)\label{eq:Hamiltoniam_triple}
\end{equation}
where $r$ represents the position of mass $m_{1}$ relative to mass
$m_{0}$, $r_{02}$ represents the position of mass $m_{2}$ relative
to mass $m_{0}$ and $r_{12}$ represents the position of mass $m_{1}$
relative to mass $m_{2}$. The first term is the kinetic energy of
the inner orbit and the second term is the kinetic energy of the outer
orbit. The last term is the potential energy of the triple system
\begin{equation}
V\left(r,r_{12},r_{23}\right)=\frac{-Gm_{0}m_{1}}{r}+\frac{-Gm_{0}m_{2}}{r_{02}}+\frac{-Gm_{1}m_{2}}{r_{12}}\label{eq:potential energy}
\end{equation}
The complete Hamiltonian, expanded in powers of $\alpha$ is then
\[
\mathcal{H}=-\frac{Gm_{0}m_{1}}{2a_{1}}-\frac{G\left(m_{0}+m_{1}\right)m_{2}}{2a_{2}}
\]
 
\begin{equation}
-\frac{G}{a_{2}}\sum_{j=2}^{\infty}\alpha^{j}M_{j}\left(\frac{r}{a_{1}}\right)^{j}\left(\frac{a_{2}}{R}\right)^{j+1}P_{j}\left(\cos\psi\right)\label{eq:Triple_Hamilotonian_heirachichal}
\end{equation}
where $\psi$ is the angle between $r$ and $R$ (see Fig.\ref{fig:Heirarchichal-theree-body})
and $P_{j}$ is the $j$-th Legendre polynomial. The first and second
terms are the total energy of the inner and outer binary respectively,
the third term is the expanded series corresponding to the interaction
between the two orbits. It is convenient to write the Hamiltonian
(\ref{eq:Triple_Hamilotonian_heirachichal}) in the angle-action variables
called \textit{Delaunay's elements} \citet{Valtonen2006} which provide
a convenient dynamical description of the three-body system. The coordinates
are chosen to be the mean anomalies, $l_{1}$ and $l_{2}$, and their
conjugate momenta
\[
L_{1}=\frac{m_{0}m_{1}}{m_{0}+m_{1}}\sqrt{G\left(m_{0}+m_{1}\right)a_{1}}
\]
 
\begin{equation}
L_{2}=\frac{m_{2}\left(m_{0}+m_{1}\right)}{m_{0}+m_{1}+m_{3}}\sqrt{G\left(m_{0}+m_{1}+m_{2}\right)a_{2}},
\end{equation}
as well as the arguments of the periastron , $g_{1}$ and $g_{2}$,
and their conjugate momenta (the orbital angular momenta of the orbits),
\begin{equation}
G_{1}=L_{1}\sqrt{1-e_{1}^{2}}\qquad G_{2}=L_{2}\sqrt{1-e_{2}^{2}}\label{eq:G1-G2-definition}
\end{equation}
where $e_{1}$ and $e_{2}$ are the inner and outer orbit eccentricity,
respectively. We also make use of the longitudes of ascending nodes,
$h_{1}$and $h_{2}$, and their conjugate momenta (these are the z-components
of the orbital angular momenta of the orbits) 
\begin{equation}
H_{1}=G_{1}\cos i_{1}\qquad H_{2}=G_{2}\cos i_{2}\label{eq:H1-H2-definition}
\end{equation}
and 
\begin{equation}
H\equiv G_{tot}=G_{1}\cos i_{1}+G_{2}\cos i_{2}\label{eq:H - definition}
\end{equation}
where subscripts $1,2$ denote the inner and outer orbits, respectively
and $H$ is the total angular momentum. For geometric intuition see
Fig. \ref{fig:Delaunays elements}.
 Using this coordinate system the Hamiltonian can be written in third
order in $\alpha$ (octupole approximation) \citet{Naoz2013,Blaes2002}.
\[
\mathcal{H}=-\frac{\beta_{0}}{2L_{1}^{2}}-\frac{\beta_{1}}{2L_{2}^{2}}-4\beta_{2}\left(\frac{L_{1}^{4}}{L_{2}^{6}}\right)\left(\frac{r_{1}}{a_{1}}\right)^{2}\left(\frac{a_{2}}{r_{2}}\right)^{3}\left(3\cos2\psi+1\right)
\]
 
\begin{equation}
-2\beta_{3}\left(\frac{L_{1}^{6}}{L_{2}^{8}}\right)\left(\frac{r_{1}}{a_{1}}\right)^{3}\left(\frac{a_{2}}{r_{2}}\right)^{4}\left(5\cos^{3}\psi-3\cos\psi\right)\label{eq:ouctopole hamiltonian}
\end{equation}
where 
\begin{equation}
\beta_{0}=Gm_{0}m_{1}\frac{L_{1}^{2}}{a_{1}}
\end{equation}
\begin{equation}
\beta_{1}=Gm_{2}\left(m_{0}+m_{1}\right)\frac{L_{2}^{2}}{a_{2}}
\end{equation}
and 
\begin{equation}
\beta_{2}=\frac{G^{2}}{16}\frac{\left(m_{0}+m_{1}\right)^{7}m_{2}^{7}}{\left(m_{0}m_{1}\right)^{3}\left(m_{0}+m_{1}+m_{2}\right)^{3}}
\end{equation}
\begin{equation}
\beta_{3}=\frac{G^{2}}{4}\frac{\left(m_{0}+m_{1}\right)^{9}m_{2}^{9}\left(m_{0}-m_{1}\right)}{\left(m_{0}m_{1}\right)^{5}\left(m_{0}+m_{1}+m_{2}\right)^{4}}
\end{equation}

In order to get the equations of motion one should use the Hamilton
equation with respect to the relevant Hamiltonian. However, in the
case of hierarchical systems a separation of timescales approach can
be employed. The short timescale is the inner and outer orbit period,
the long timescale is the timescale for the change in the orbital
elements. The timescale difference enables us to average over the
short timescales and obtain a much simpler Hamiltonian, and hence
simpler set of the equations of motion. This is explained in details
in the following subsection.

\subsection{Secular Dynamics}

\label{sub:Secular-Dynamics}The Hamiltonian (\ref{eq:ouctopole hamiltonian})
contains information about the long and short time behavior of the
system. Because of the separation of timescales, one can average over
the short time evolution. This is done by averaging over rapidly varying
$l_{1}$ and $l_{2}$ in the Hamiltonian; so called double averaging,
and using the Von Zeipel method, i.e. transforming the coordinates
into different angle-action variables that align the total angular
momentum vector to the z-axis (see \citealt{Naoz2011}). Using this
method we eliminate the nodes by the relation $h_{1}-h_{2}=\pi$.
Next we define 
\begin{equation}
\theta\equiv\cos i=\frac{H^{2}-G_{1}^{2}-G_{2}^{2}}{2G_{1}G_{2}}\label{eq:theta}
\end{equation}
where $i=i_{1}+i_{2}$, is the total inclination (see Fig. (\ref{fig:Delaunays elements})).
One then obtains the equations of motion through the canonical equations
(\citealt{Ford2000,Blaes2002,Naoz2011}):
\[
\frac{dg_{1}}{dt}=6C_{2}\left\{ \frac{1}{G_{1}}[4\theta^{2}+\left(5\cos2g_{1}-1\right)\left(1-e_{1}^{2}-\theta^{2}\right)\right\} 
\]
 
\[
+6C_{2}\left\{ \frac{\theta}{G_{2}}\left[2+e_{1}^{2}\left(3-5\cos2g_{1}\right)\right]\right\} 
\]
\[
+C_{3}e_{2}e_{1}\left(\frac{1}{G_{2}}+\frac{\theta}{G_{1}}\right)\left\{ \sin g_{1}\sin g_{2}\left[A+10\left(3\theta^{2}-1\right)\left(1-e_{1}^{2}\right)\right]\right\} 
\]
\[
-C_{3}e_{2}e_{1}\left(\frac{1}{G_{2}}+\frac{\theta}{G_{1}}\right)\left\{ 5\theta B\cos\psi\right\} 
\]
\[
-C_{3}e_{2}\frac{1-e_{1}^{2}}{e_{1}G_{1}}\left[10\theta\left(1-\theta^{2}\right)\left(1-3e_{1}^{2}\right)\sin g_{1}\sin g_{2}\right]
\]
\begin{equation}
-C_{3}e_{2}\frac{1-e_{1}^{2}}{e_{1}G_{1}}\left[\cos\phi\left(3A-10\theta^{2}+2\right)\right]\label{eq: g1-EOM}
\end{equation}
\\
\[
\frac{de_{1}}{dt}=30C_{2}\frac{e_{1}\left(1-e_{1}^{2}\right)}{G_{1}}\left(1-\theta^{2}\right)\sin2g_{1}-C_{3}e_{2}\frac{1-e_{1}^{2}}{G_{1}}\times
\]
 
\[
\left[35\cos\psi\left(1-\theta^{2}\right)e_{1}^{2}\sin2g_{1}\right]-C_{3}e_{2}\frac{1-e_{1}^{2}}{G_{1}}\times
\]
\[
\left[-10\theta\left(1-e_{1}^{2}\right)\left(1-\theta^{2}\right)\cos g_{1}\sin g_{2}\right]-C_{3}e_{2}\frac{1-e_{1}^{2}}{G_{1}}\times
\]
\begin{equation}
\left[-A\left(\sin g_{1}\cos g_{2}-\theta\cos g_{1}\sin g_{2}\right)\right]\label{eq:e1-EOM}
\end{equation}
\\
\[
\frac{dg_{2}}{dt}=3C_{2}\left\{ \frac{2\theta}{G_{1}}\left[2+e_{1}^{2}\left(3-5\cos2g_{1}\right)\right]\right\} 
\]
\[
+3C_{2}\left\{ \frac{1}{G_{2}}\left[4+6e_{1}^{2}+\left(5\theta^{2}-3\right)\left(2+3e_{1}^{2}-5e_{1}^{2}\cos2g_{1}\right)\right]\right\} 
\]
\[
-C_{3}e_{1}\sin g_{1}\sin g_{2}\left\{ \frac{4e_{2}^{2}+1}{e_{2}G_{2}}10\theta\left(1-\theta^{2}\right)\left(1-e_{1}^{2}\right)\right\} 
\]
\[
-C_{3}e_{1}\sin g_{1}\sin g_{2}\left\{ -e_{2}\left(\frac{1}{G_{1}}+\frac{\theta}{G_{2}}\right)\right.
\]

\[
\left.\times\left[A+10\left(3\theta^{2}-1\right)\left(1-e_{1}^{2}\right)\right]\right\} 
\]

\begin{equation}
-C_{3}e_{1}\cos\psi\left[5B\theta e_{2}\left(\frac{1}{G_{1}}+\frac{\theta}{G_{2}}\right)+\frac{4e_{2}^{2}+1}{e_{2}G_{2}}A\right]\label{eq:g2-EOM}
\end{equation}
\\
\[
\frac{de_{2}}{dt}=C_{3}e_{1}\frac{1-e_{2}^{2}}{G_{2}}\left[10\theta\left(1-\theta^{2}\right)\left(1-e_{1}^{2}\right)\sin g_{1}\cos g_{2}\right]
\]
\begin{equation}
+C_{3}e_{1}\frac{1-e_{2}^{2}}{G_{2}}\left[A\left(\cos g_{1}\sin g_{2}-\theta\sin g_{1}\cos g_{2}\right)\right]\label{eq:e2-EOM}
\end{equation}

\begin{equation}
\frac{dG_{1}}{dt}=-C_{2}\cdot30e_{1}^{2}\sin2g_{1}\left(1-\theta^{2}\right)\label{eq:G1-EOM}
\end{equation}

\[
-C_{3}e_{1}e_{2}\left(-35e_{1}^{2}\left(1-\theta^{2}\right)\sin2g_{1}\cos\psi\right)
\]
\[
-C_{3}e_{1}e_{2}\left(A\left(\sin g_{1}\cos g_{2}-\theta\cos g_{1}\sin g_{2}\right)\right)
\]
\[
-C_{3}e_{1}e_{2}\left(10\theta\left(1-\theta^{2}\right)\left(1-e_{1}^{2}\right)\cos g_{1}\sin g_{2}\right)
\]
\[
\frac{dG_{2}}{dt}=-C_{3}e_{1}e_{2}\cdot A\left(\cos g_{1}\sin g_{2}-\theta\sin g_{1}\cos g_{2}\right)
\]
\begin{equation}
-C_{3}e_{1}e_{2}\cdot10\theta\left(1-\theta^{2}\right)\left(1-e_{1}^{2}\right)\sin g_{1}\cos g_{2}.\label{eq:G2-EOM}
\end{equation}
from Eq. (\ref{fig:Delaunays elements}), the law of sine and the
geometrical relations 
\begin{equation}
H_{1}=\frac{H^{2}+G_{1}^{2}-G_{2}^{2}}{2H}\label{eq:H1-geometry}
\end{equation}
and 
\begin{equation}
H_{2}=\frac{H^{2}+G_{2}^{2}-G_{1}^{2}}{2H}\label{eq:H2-geometry}
\end{equation}
one can get the z-component angular momentum equation of motion 
\begin{equation}
\frac{dH_{1}}{dt}=\frac{\sin i_{2}}{\sin i_{tot}}\dot{G}_{1}-\frac{\sin i_{1}}{\sin i_{tot}}\dot{G}_{2}\label{eq:H1-EOM}
\end{equation}
\begin{equation}
\frac{dH_{2}}{dt}=\frac{\sin i_{1}}{\sin i_{tot}}\dot{G}_{2}-\frac{\sin i_{2}}{\sin i_{tot}}\dot{G}_{1}.\label{eq:H2-EOM}
\end{equation}
Now, using the geometrical relations $\cos i_{1}=\left(H^{2}+G_{1}^{2}+G_{2}^{2}\right)/2HG_{1}$
and $\cos i_{2}=\left(H^{2}+G_{1}^{2}+G_{2}^{2}\right)/2HG_{2}$ we
obtain 
\begin{equation}
\frac{d\cos i_{1}}{dt}=\frac{\dot{H}_{1}}{G_{1}}-\frac{\dot{G}_{1}\cos i_{1}}{G1}\label{eq:cosi1-EOM}
\end{equation}
\begin{equation}
\frac{d\cos i_{2}}{dt}=\frac{\dot{H}_{2}}{G_{2}}-\frac{\dot{G}_{2}\cos i_{2}}{G_{2}}\label{eq:cosi2-EOM}
\end{equation}
where 
\begin{equation}
C_{2}=\frac{Gm_{0}m_{1}m_{2}}{16\left(m_{0}+m_{1}\right)a_{2}\left(1-e_{2}^{2}\right)^{3/2}}\left(\frac{a_{1}}{a_{2}}\right)^{2}\label{eq:C2-definition}
\end{equation}
\begin{equation}
C_{3}=\frac{15Gm_{0}m_{1}m_{2}\left(m_{2}-m_{1}\right)}{64\left(m_{0}+m_{1}\right)^{2}a_{2}\left(1-e_{2}^{2}\right)^{5/2}}\left(\frac{a_{1}}{a_{2}}\right)^{3}.\label{eq:C3-definition}
\end{equation}
are the quadruple and octupole coefficients respectively. The quantities
$A$ and $B$ in these terms are given by 
\begin{equation}
A=4+e_{1}^{2}-\frac{5}{2}\left(1-\theta^{2}\right)B\label{eq:A defintion}
\end{equation}
and 
\begin{equation}
B=2+5e_{1}^{2}-7e_{1}^{2}\cos2g_{1}\label{eq:B defintion}
\end{equation}
and $\psi$ is the angle between the periastron directions, 
\begin{equation}
\cos\psi=-\cos g_{1}\cos g_{2}-\theta\sin g_{1}\sin g_{2}.\label{eq:cos_phi definition}
\end{equation}
The $C_{3}$ then makes it clear that the octupole terms vanish whence
$m_{0}=m_{1}$. 

Let us now define 
\begin{equation}
\epsilon_{3}=\left(\frac{m_{0}-m_{1}}{m_{0}+m_{1}}\right)\left(\frac{a_{1}}{a_{2}}\right)\frac{e_{2}}{1-e_{2}^{2}}\label{eq:importance of octupole}
\end{equation}
which measures the importance of the octupole terms, $C_{3}/C_{2}$.
Note that in this approximation the energy for the inner and outer
orbits are conserved, respectively; namely $\dot{a}_{1}=0$ and $\dot{a}_{2}=0$,
and similarly the total angular momentum $\dot{H}=0$. Hence, the
interaction only involves exchanging angular momentum between the
two orbits.

We obtain the equations of motion of the orbital elements. As can
be seen the secular evolution, as described by the quadruple level
equations, gives rise to an oscillatory behavior, where both the eccentricity
and the inclination of the inner binary periodically change with a
potentially large amplitude; these are termed Kozai-Lidov oscillations/cycles
\citep{Kozai1962,Lidov1962} which have been discussed extensively
in the past recent years. The characteristic timescale of a full Kozai
cycle is given by \citep{Kozai1962,Valtonen2006} 
\begin{equation}
P_{Kozai}\approx\left(\frac{a_{2}}{a_{1}}\right)^{3}\frac{m_{0}+m_{1}}{m_{2}}P_{1}=\frac{P_{2}^{2}}{P_{1}}\left(\frac{m_{0}+m_{1}}{m_{2}}\right)\label{eq:Kozai-period}
\end{equation}
where 
\begin{equation}
P_{1}=2\pi\sqrt{a_{1}^{3}/G\left(m_{0}+m_{1}\right)}\label{eq: inner period}
\end{equation}
is the period time of the inner binary and 
\begin{equation}
P_{2}=2\pi\sqrt{a_{2}^{3}/G\left(m_{2}+m_{1}+m_{0}\right)}
\end{equation}
is the period time of the outer binary. The maximum inner eccentricity
during a cycle is given by \citet{Valtonen2006} 
\begin{equation}
e_{max}=\sqrt{1-\frac{5}{3}\cos^{2}i_{0}}\label{eq:maximum e}
\end{equation}
where $i_{0}$ is the initial inclination between the inner and outer
binary in the triple (assuming an initial circular orbit).

In the octupole regime, an additional, long term low frequency modulation
can be seen on a different timescale (see section \ref{sub:Timescales}).
In particular, in this regime the system evolution is much more complex,
and it is not periodic on the long term. Moreover, the amplitude of
the changes in the inclinations and eccentricities could be much larger.
In this regime transitions between prograde and retrograde orbits
due to extremely high eccentricities can also occur. The latter effect
becomes highly important when discussing realistic non-point particles,
where close encounters between the inner triple components can lead
to significant orbital changes through the coupling of short-range
dissipative processes. 

Making use of the secular formulation for the evolution of classical
point-like particles described above, we are now ready to take the
next step, and couple the triple dynamics with dissipative and non-classical
processes. The secular classical point-particle dynamics of the triple
systems described above conserve energy, and therefore the SMAs of
the system remain constant throughout their evolution, $a_{1}=const$
, $a_{2}=const$. In the following subsections general relativistic
(GR) interaction, mass loss and mass transfer processes are described.
In a realistic triple system these effects are important and influence
the equation of motion. In section \ref{sub:Post-Newtonian-terms}
we consider the general relativistic effects on the equation of motion,
and in Section \ref{sec:Secular-evolution-with-massloss} we added
our novel treatment for the mass loss and mass transfer processes.

\subsection{Post Newtonian terms}

\label{sub:Post-Newtonian-terms}General relativity plays a key role
in close compact systems. These post Newtonian effects come in two
separate manifestations, both are taken into account only for the
inner binary, GR effects on the outer binary are typically insignificant
in comparison, besides the cases of weakly hierarchical systems, where
the secular approach becomes invalid; in the following GR effects
on the external binary are neglected (for treatment of such systems
see \citealt{Naoz2013a}). 

The first post Newtonian effect included is the precession of periastron
(not a dissipative effect but rather an additional dynamical process),
and the second is gravitational wave (GW) radiation (which is a dissipative
process). For the former, Hamiltonian analytic treatment can be used
\citep{Blaes2002}, by adding the averaged post Newtonian Hamiltonian
to the averaged Hamiltonian described in section \ref{sub:Secular-Dynamics}
\[
\mathcal{H}=C_{2}\left[\left(2+e_{1}^{2}\right)\left(1-3\theta^{2}\right)-15e_{1}^{2}\left(1-\theta^{2}\right)\cos2g_{1}\right]
\]
\[
+C_{3}e_{1}e_{2}\left[A\cos\phi+10\theta\left(1-\theta^{2}\right)\left(1-e_{1}^{2}\right)\sin g_{1}\sin g_{2}\right]
\]
\begin{equation}
+\frac{G^{2}m_{0}m_{1}}{c^{2}a_{1}^{2}}\left[\frac{15m_{1}^{2}+15m_{0}^{2}+29m_{0}m_{1}}{8\left(m_{0}+m_{1}\right)}-\frac{3\left(m_{0}+m_{1}\right)}{\left(1-e_{1}^{2}\right)^{1/2}}\right].\label{eq:Post Newtonian Hamltonian}
\end{equation}
From this Hamiltonian one obtains the additional term for the precession
of the inner binary
\begin{equation}
\frac{dg_{1}}{dt}=\frac{3}{c^{2}a_{1}\left(1-e_{1}^{2}\right)}\cdot\left(\frac{G\left(m_{0}+m_{1}\right)}{a_{1}}\right)^{3/2}.\label{eq:g1 - correction}
\end{equation}
For the gravitational wave treatment one can follow \citet{Peters1964}
and compute the loss of energy, angular momentum and the change in
eccentricity, averaged per orbit, and add the appropriate terms to
the Hamiltonian equations. The GR terms are as follows; for the inner
binary SMA
\begin{equation}
\frac{da_{1}}{dt}=-\frac{64G^{3}m_{0}m_{1}\left(m_{0}+m_{1}\right)}{5c^{5}a_{1}^{3}\left(1-e_{1}^{2}\right)^{7/2}}\left(1+\frac{73}{24}e_{1}^{2}+\frac{37}{96}e_{1}^{4}\right)\label{eq:a1 -correction}
\end{equation}
for the inner orbit eccentricity
\begin{equation}
\frac{de_{1}}{dt}=-\frac{304G^{3}m_{0}m_{1}\left(m_{0}+m_{1}\right)e_{1}}{15c^{5}a_{1}^{4}\left(1-e_{1}^{2}\right)^{5/2}}\left(1+\frac{121}{304}e_{1}^{2}\right)\label{eq:e1 - correction}
\end{equation}
and for the total loss of angular momentum due to GW radiation from
the inner binary,
\[
\frac{dH}{dt}=-\frac{32G^{3}m_{0}^{2}m_{1}^{2}}{5c^{5}a_{1}^{3}\left(1-e_{1}^{2}\right)^{2}}\left[\frac{G\left(m_{0}+m_{1}\right)}{a_{1}}\right]^{1/2}\times
\]
\begin{equation}
\left(1+\frac{7}{8}e_{1}^{2}\right)\frac{G_{1}+G_{2}\theta}{H}.\label{eq:H  - correction}
\end{equation}
Note that the inner binary radiates GWs and changes the magnitude
of $G_{1}$ and $H$ while the magnitude of $G_{2}$ remains unchanged.

\subsection{Timescales}

\label{sub:Timescales}In this subsection we present the different
timescales of the dynamical problem at hand. The short time scales
of the system are the inner and outer binary periods, $P_{1}$ and
$P_{2}$, respectively. Using the double averaging method we effectively
assume no variation occurs on these (or shorter) time scales. The
next timescale corresponds to the quadrupole term, which gives rise
to the ``standard'' Kozai cycling. Its characteristic time scale
is $t_{2}\sim G_{1}/C_{2}$. More specifically, the timescale is \citep{Naoz2013a}
\begin{equation}
P_{Kozai}=t_{2}\sim2\pi\frac{a_{2}^{3}\left(1-e_{2}^{2}\right)^{3/2}\left(m_{0}+m_{1}\right)^{1/2}}{G^{1/2}m_{2}a_{1}^{3/2}}.\label{eq:quad_time scale}
\end{equation}

The timescale corresponding to the octupole level perturbations has
the form of $t_{3}\sim\epsilon_{3}t_{2}$ and the explicit term is
\begin{equation}
t_{3}\sim2\pi\frac{a_{2}^{4}\left(1-e_{2}^{2}\right)^{5/2}\left(1-e_{1}^{2}\right)^{1/2}\left(m_{0}+m_{1}\right)^{3/2}}{G^{1/2}m_{2}\left|m_{0}-m_{1}\right|e_{2}a_{1}^{5/2}}.\label{eq:octupole_time}
\end{equation}
The long term modulation of the standard Kozai cycles occurs on this
timescale. 

An additional important timescale is that arising from GR effects,
the time scales corresponding to the different post-Newtonian terms,
which has two manifestations. The first is the precession time scale
of the inner binary, and the second it the GW radiation time scale
from the inner binary. For the GR precession time scale 
\begin{equation}
t_{1PN}\sim2\pi\frac{a_{1}^{5/2}c^{2}\left(1-e_{1}^{2}\right)}{3G^{3/2}\left(m_{0}+m_{1}\right)^{3/2}}\label{eq:percession timescale}
\end{equation}
 which corresponds to the time the ellipse makes a complete revolution
due to GR precession ($1PN$ stands for the first Post-Newtonian term).
The GW timescale is 
\begin{equation}
t_{2.5PN}\sim\frac{c^{5}a_{1}^{4}\left(1-e_{1}^{2}\right)^{7/2}}{G^{3}m_{0}m_{1}\left(m_{0}+m_{1}\right)},\label{eq:GW timescale}
\end{equation}
where $2.5PN$ stands for the $2.5$ Post-Newtonian term. The physical
meaning of this post Newtonian term is the timescale for orbital energy
loss due to GW emission, which could, eventually lead to the inner
binary merger. 

Equipped with all the necessary equations arising from the gravitational
dynamics, we now continue to add relevant terms arising from the evolutionary
processes of mass-loss and mass-transfer to the equations of motion.

\section{Secular evolution with mass loss and mass transfer}

\label{sec:Secular-evolution-with-massloss}In this section we treat
the secular evolution equations of motion with mass loss. For a single
star, mass loss rate vary on a wide range, from $\dot{M}\sim10^{-14}M_{\odot}yr^{-1}$
to $\dot{M}\sim1M_{\odot}yr^{-1}$, depending on stellar mass, evolutionary
stage and the mass loss mechanism, e.g. a supernova (SN) explosion
causes prompt mass loss, and a common envelope (CM) stage in a binary
system expel up to a few solar mass per year. Mass loss is therefore
an important evolutionary effect for single, binary and triple stellar
evolution. Here we treat only secular mass-loss/mass-transfer processes,
which are defined as slow changes in mass, with respect to the orbital
period, namely 
\begin{equation}
P_{1}\cdot\dot{M_{b}}\ll M_{b}\label{eq:adiabatic condition}
\end{equation}
where $M_{b}$ is the total mass of a given mass-losing binary system.
For simplicity we assume secular mass loss of mass $m_{0}$ in the
inner binary of the form 
\begin{equation}
\frac{d}{dt}m_{0}=-\alpha
\end{equation}
where $\alpha$ is an arbitrary function satisfying (\ref{eq:adiabatic condition}).
The corrected equations of motion are presented for this case; a complete
treatment for mass loss from both the inner and outer binary is presented
the end of this section. For secular mass loss in a binary, the SMA
and the eccentricity evolve according the following equations of motion
\begin{equation}
\dot{a}_{ML}=-\frac{\frac{d}{dt}\left(m_{0}+m_{1}\right)}{\left(m_{0}+m_{1}\right)}a\label{eq:SMA_M_DOT}
\end{equation}
\begin{equation}
\dot{e}_{ML}=0
\end{equation}
where ML refers to mass loss \citet{Gurfil2008,Hadjidemetriou1963,Shappee2013,Veras2011}.
This corresponds to an additional term in (\ref{eq:a1 -correction})
\begin{equation}
\dot{a}_{1,ML}=\frac{\alpha}{m_{0}+m_{1}}a_{1}
\end{equation}
and a similar change in $a_{2}$, which was originally a constant
of motion, in the absence of mass loss processes
\begin{equation}
\dot{a}_{2,ML}=\frac{\alpha}{m_{2}+\left(m_{0}+m_{1}\right)}a_{2}.
\end{equation}

The rate of change of the orbital angular momentum $G_{1}$ with mass
loss is given by differentiating (\ref{eq:G1-G2-definition}) with
respect to time 
\begin{equation}
\dot{G}_{1,ML}=\frac{-\alpha m_{1}}{m_{0}\left(m_{0}+m_{1}\right)}G_{1}
\end{equation}
and the outer angular momentum $G_{2}$ follows 
\begin{equation}
\dot{G}_{2,ML}=\frac{-\alpha m_{2}}{\left(m_{0}+m_{1}\right)\left(m_{0}+m_{1}+m_{2}\right)}G_{2}.\label{eq:G2_ML}
\end{equation}
These additional terms change only in magnitude and not in direction.
Note the angular momentum \emph{per unit mass} is conserved; hence
at the \emph{test particle limit}, the angular momentum itself would
also be conserved. The total angular momentum change within one orbital
period time, due to the mass loss in the inner binary, is given by
differentiating (\ref{eq:H - definition}) \textbf{
\begin{equation}
\dot{H}_{ML}=\dot{G}_{1}\cos i_{1}+\dot{G}_{2}\cos i_{2}.
\end{equation}
}The change in the total angular momentum $H$ effects (\ref{eq:H1-EOM})
and (\ref{eq:H2-EOM}). Therefore one needs to recalculate the rate
of change of $H_{1}$ and $H_{2}$. Taking the time derivatives of
Eqs. (\ref{eq:H1-geometry}) and (\ref{eq:H2-geometry}) adds an additional
term proportional to $\dot{H}$
\begin{equation}
\dot{H}_{1,ML}=\dot{H}\left(1-\frac{H_{1}}{H}\right)
\end{equation}
 
\begin{equation}
\dot{H}_{2,ML}=\dot{H}\left(1-\frac{H_{2}}{H}\right).
\end{equation}
In the general case where all three objects losses mass 
\begin{equation}
\frac{d}{dt}m_{2}=-\beta
\end{equation}
\begin{equation}
\frac{d}{dt}m_{1}=-\gamma
\end{equation}
where $\beta$ and $\gamma$ are arbitrary functions satisfying (\ref{eq:adiabatic condition}),
the additional terms added to the inner and outer angular momentum
(\ref{eq:G1-EOM}), (\ref{eq:G2_ML}) are
\begin{equation}
\dot{G}_{1,ML}=\frac{-\alpha m_{1}}{m_{0}\left(m_{0}+m_{1}\right)}G_{1}+\frac{-\gamma m_{0}}{m_{1}\left(m_{0}+m_{1}\right)}G_{1}
\end{equation}
 
\[\dot{G}_{2,ML}=\frac{G_2(\alpha-\gamma)m_{2}}{\left(m_{0}+m_{1}\right)\left(m_{0}+m_{1}+m_{2}\right)}
\]

\begin{equation}
+\frac{-G_2\beta\left(m_{0}+m_{1}\right)}{m_{2}\left(m_{0}+m_{1}+m_{2}\right)}
\end{equation}
and the changes in the SMAs of the inner and outer orbits are:
\begin{equation}
\dot{a}_{1,ML}=\frac{\alpha+\gamma}{m_{0}+m_{1}}a_{1}
\end{equation}
 
\begin{equation}
\dot{a}_{2}=\frac{\alpha+\gamma+\beta}{m_{2}+\left(m_{0}+m_{1}\right)}a_{2}.
\end{equation}
The functions $\alpha,\beta$ and $\gamma$ usually obtained from
stellar evolution codes or analytic approximations. 

In order to account for mass-transfer between the triple components,
we also need to introduce efficiency parameters $\psi_{0,1}$, $\psi_{1,0}$
and $\psi_{2,01}$. These correspond to the efficiency of mass transfer
in a given binary sub-system in the triple: $\psi_{0,1}$ is the efficiency
of the mass transfer from $m_{0}$ to $m_{1},$$\psi_{1,0}$ is the
efficiency of the mass transfer from $m_{1}$ to $m_{0}$ and $\psi_{2,01}$
is the efficiency of the mass transfer from $m_{2}$ to the inner
binary. It is important to note that throughout our modeling of the
mass transfer we assume no angular momentum exchange, beside that
arising directly from the mass changes; i.e. we assume mass ``disappears''
from one object and ``reappears'' in another. We can then obtain
the following general set of equations which can also account for
mass transfer 
\begin{equation}
\dot{m}_{0}=-\alpha+\psi_{1,0}\cdot\gamma+\psi_{2,01}\cdot\beta\cdot\frac{m_{0}}{m_{0}+m_{1}}\label{eq:m0_dot_mass_transfer}
\end{equation}
\begin{equation}
\dot{m}_{1}=-\gamma+\psi_{0,1}\cdot\alpha+\psi_{2,01}\cdot\beta\cdot\frac{m_{1}}{m_{0}+m_{1}}.\label{eq:m1_dot_mass_trnasfer}
\end{equation}
Note that the last term in Eqs. (\ref{eq:m0_dot_mass_transfer}) and
(\ref{eq:m1_dot_mass_trnasfer}) is obtained from mass transfer from
the tertiary companion, with the assumption that the mass accretion
is higher into the more massive object in the inner binary (following
the results by\citealp{deVries2014}).

\section{Evolutionary Channels}

\label{sec:Evolutionary-Channels}In sections \ref{sub:Timescales}
and \ref{sub:Secular-Dynamics} we presented the relevant timescales,
and the parameter determining the importance of the octupole term
in the secular dynamics equations of motion. These timescales are
dependent on the masses and SMAs of the triple components. By varying
the mass we change the relevant timescale and the dynamics, as well
as the relative importance of the octupole term. Mass-loss/mass-transfer
can therefore change the dynamical behavior of a triple system from
one regime of evolution to another. In this section we provide examples
of several triple systems showing such changes, by fully integrating
the equations of motions for these systems. In addition, we use direct
N-body integration (based on \citealp{Hut1981}) and compare them
to the results from our secular evolution method. In section \ref{sub:Mass-loss-from-inner}
we present a case of mass loss in inner binary, followed by the case
of mass loss from the the third outer companion in section \ref{sub:Mass-loss/transfer-from}.

\subsection{Mass loss from the primary components in the inner binary, and the
mass loss induced eccentric Kozai (MIEK) process}

\label{sub:Mass-loss-from-inner}\citet{Shappee2013} studied the
case of mass-loss from a component in the inner binary, which leads
to a transition from a more regular Kozai-Lidov secular behavior to
the regime where octupole level perturbations become significant,
and the amplitude of eccentricity changes become significant; a behavior
they termed mass-loss induced eccentric Kozai (MIEK). They used a
full N-body integration, which provides an excellent test case for
comparison with out secular evolution approach. 

\begin{figure*}[t]
\includegraphics[scale=0.4]{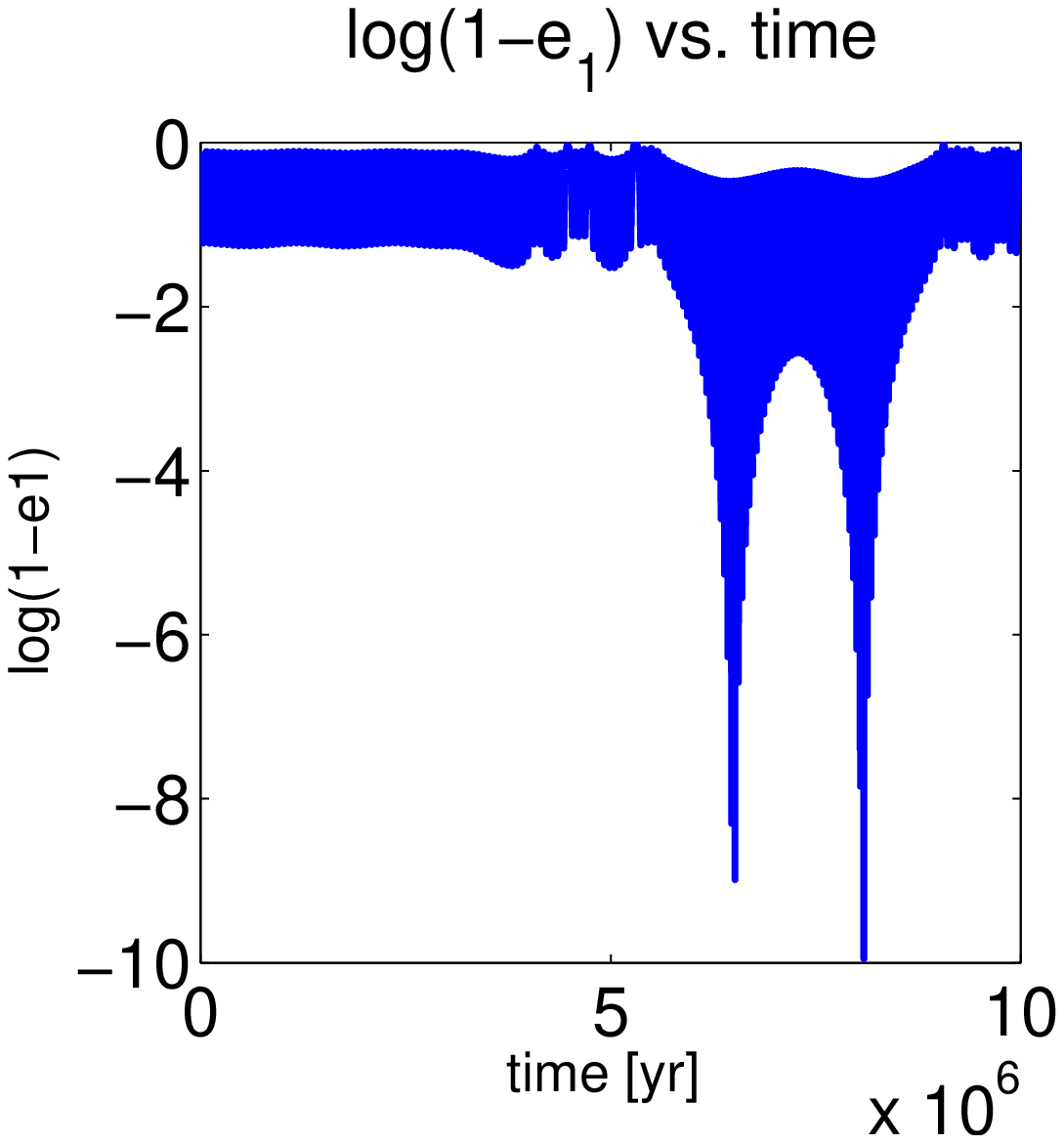}\includegraphics[scale=0.4]{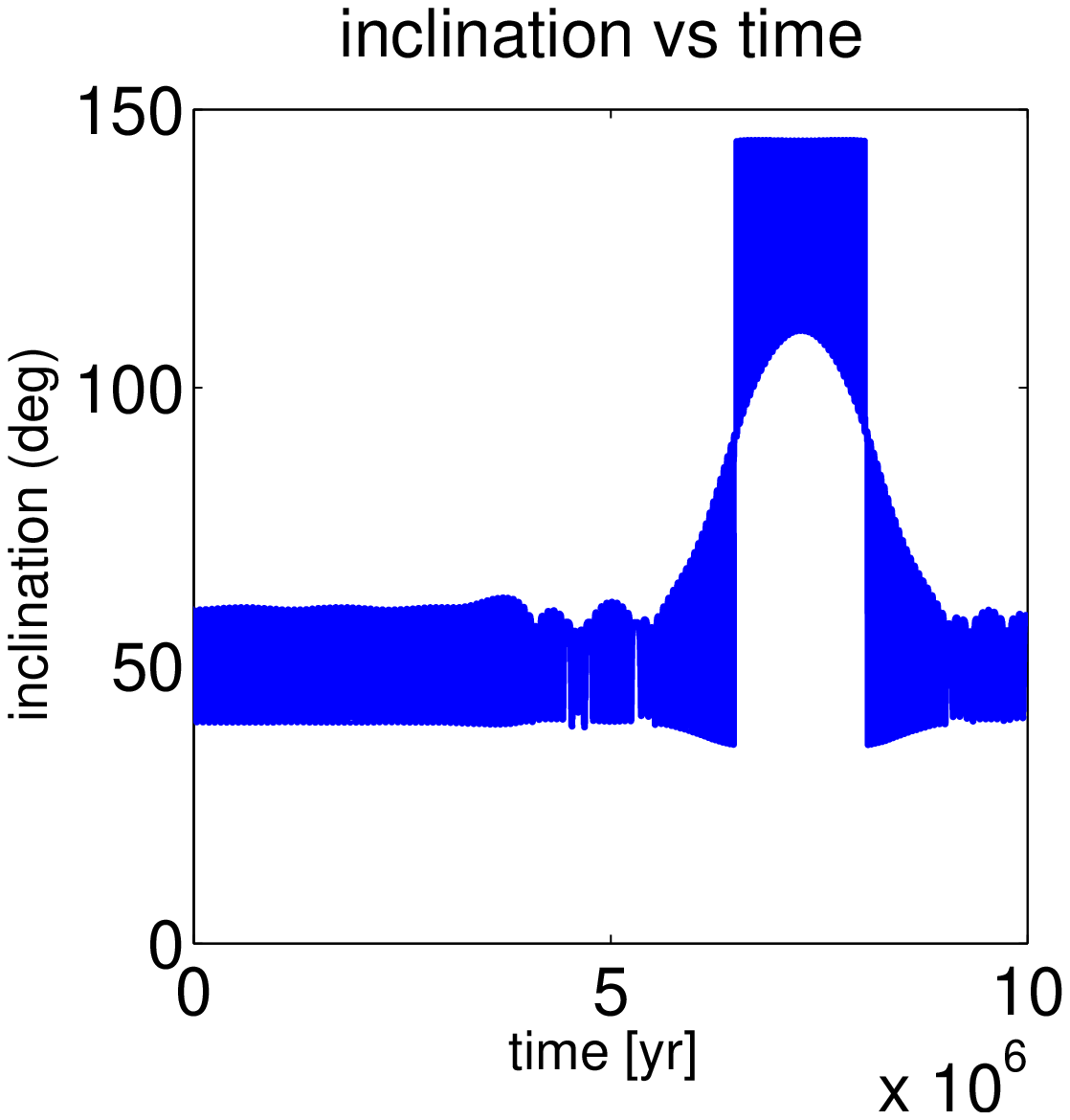}

\includegraphics[scale=0.4]{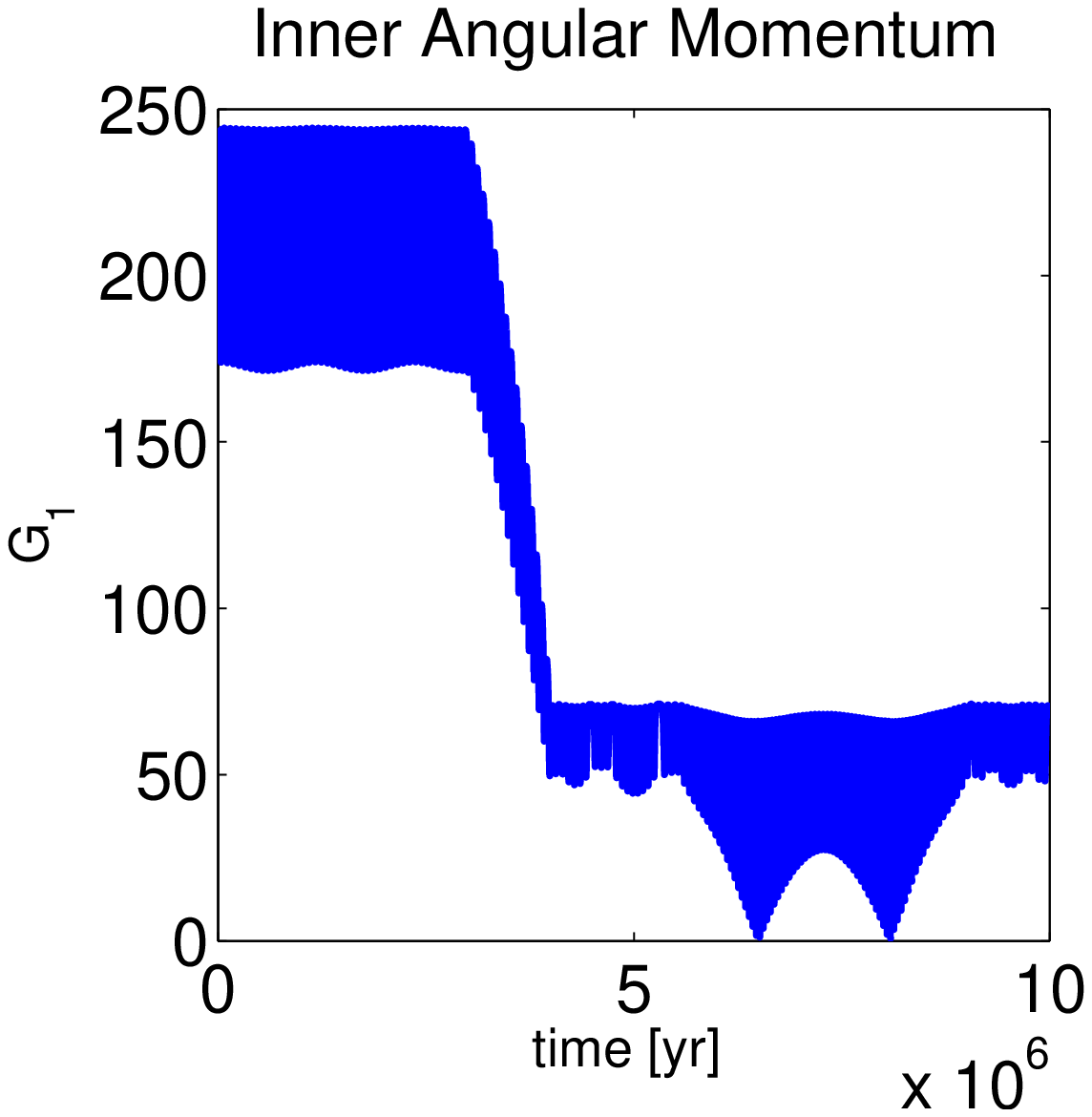}\includegraphics[scale=0.4]{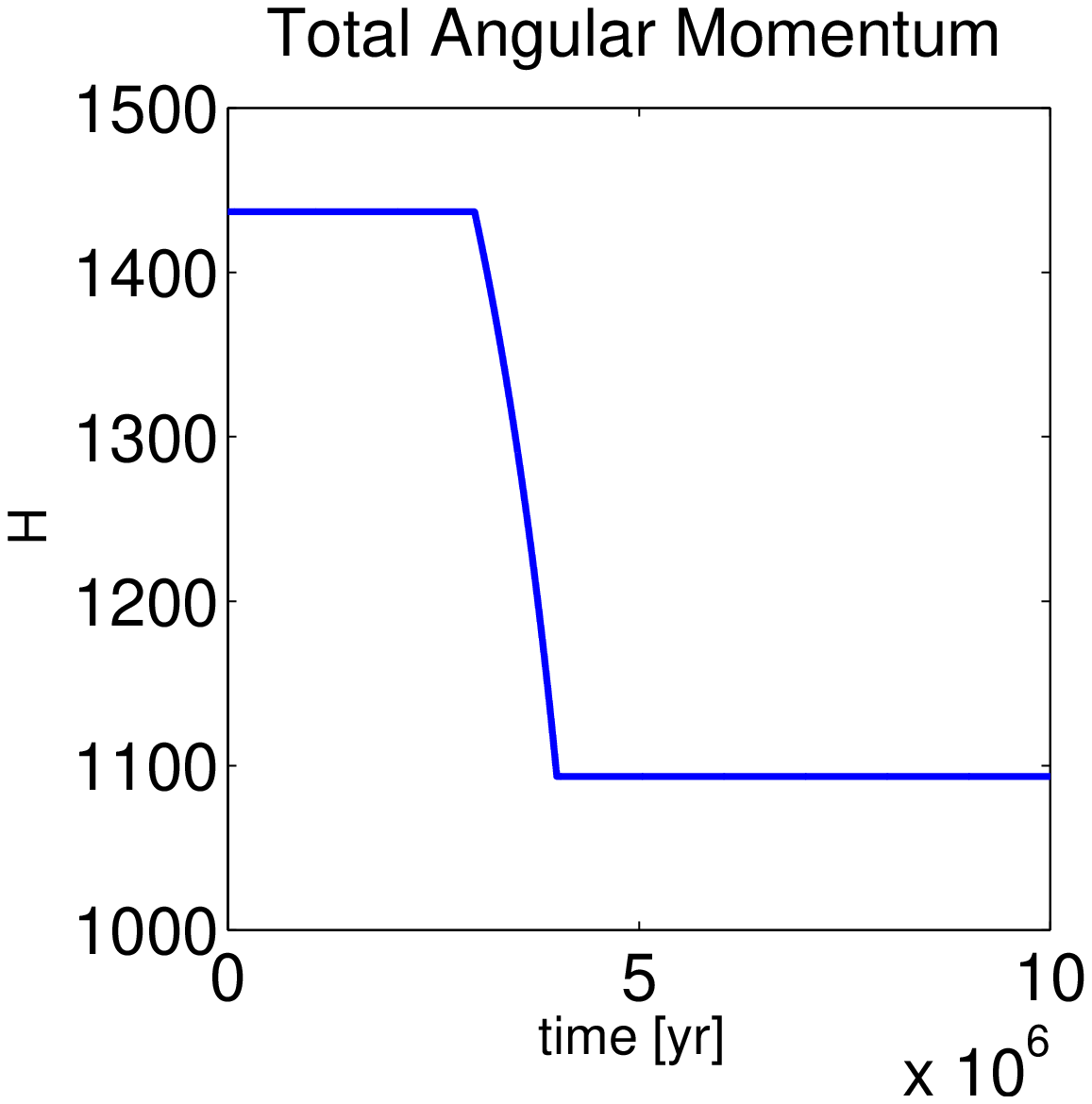}\caption{\label{fig:MIEK} Evolution of a triple system with mass loss from
the inner binary primary component, showing a MIEK-like behavior.
The initial setup of the system is given by $m_{0}=7.0\, M_{\odot}$,
$m_{1}=6.5\, M_{\odot}$, $m_{2}=6\, M_{\odot}$, $a_{1}=10\: AU$,
$a_{2}=250\, AU$, $e_{1}=0.1$, $e_{2}=0.7$, $g_{1}=0^{\circ}$,
$g_{2}=0^{\circ}$ and the mutual inclination is $i=60^{\circ}$,
similar to the case studied by \citet{Shappee2013}. Constant mass
loss is introduced for $m_{0}$ after $t=3\, Myr$ for $\Delta t_{ml}=1\, Myr$
until $m_{0}\left(t=4\, Myr\right)=1.15\, M_{\odot}$. No mass transfer
is considered. Top right: the evolution of the inclination with time.
Top left: the evolution of the inner binary eccentricity with time.
Bottom right: the evolution of the inner angular momentum, $G_{1}\left(t\right)$.
Bottom left: the evolution of the total angular momentum, $H\left(t\right)$.
After mass loss, from $t=4\, Myr$, the system enters the octupole
regime and display octupole level of evulotion,i.e. high eccentric
inner binary and even retrograde orbits.}
\end{figure*}

\begin{figure*}
\includegraphics[scale=0.4]{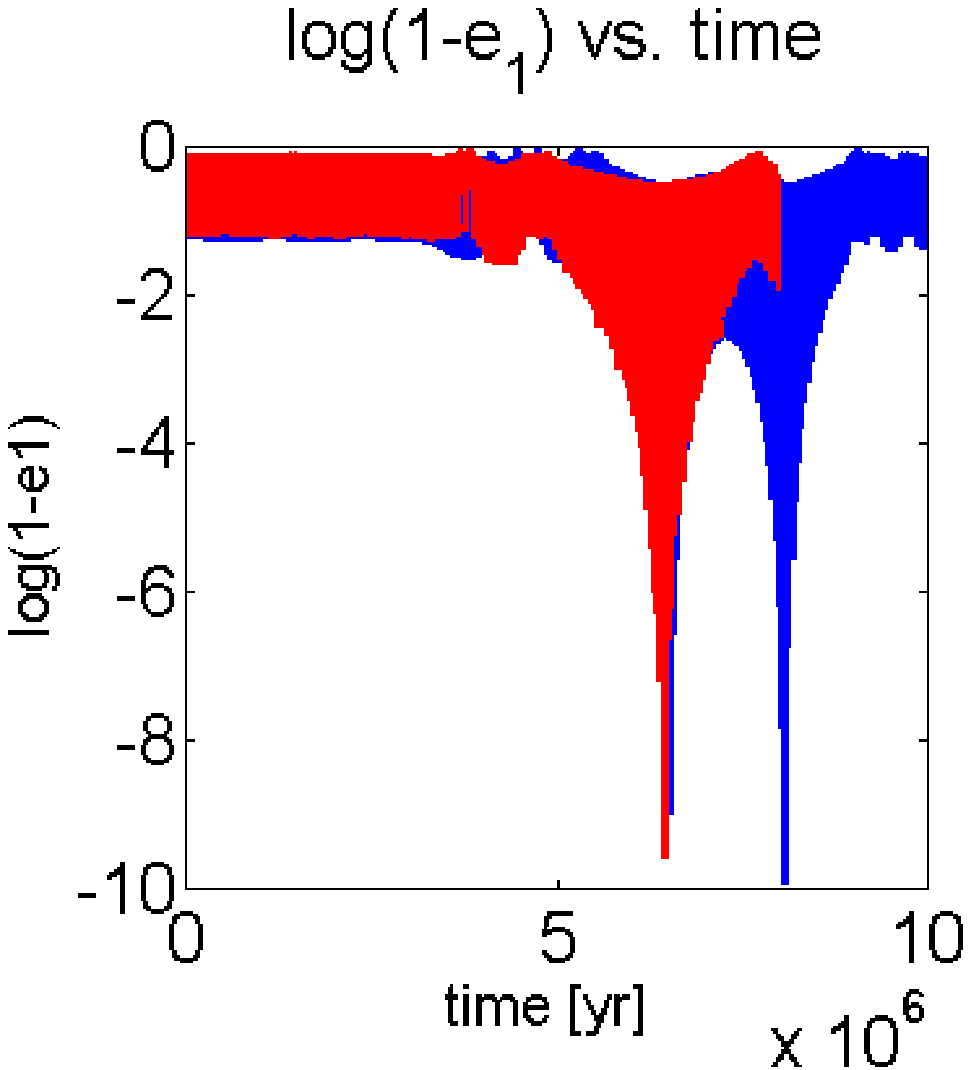}\includegraphics[scale=0.4]{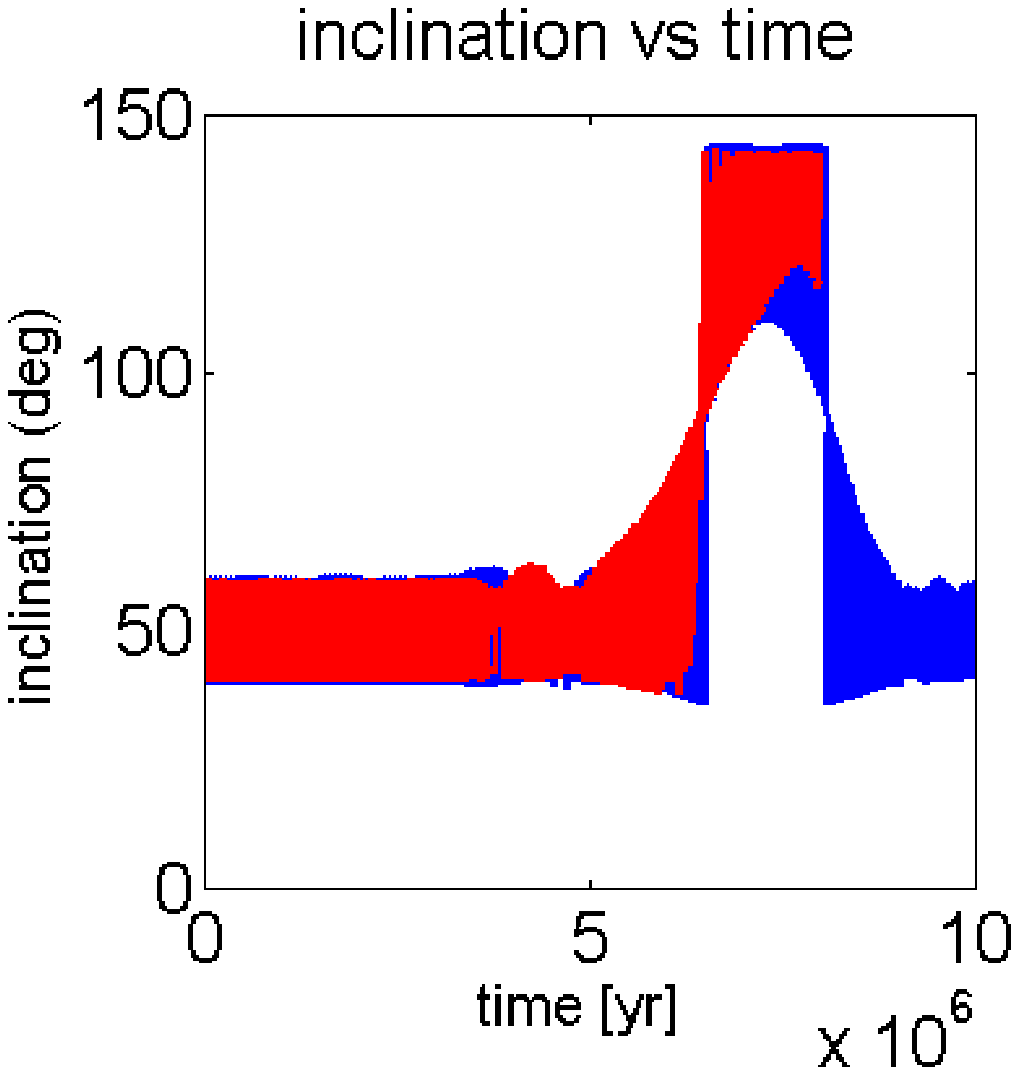}

\includegraphics[scale=0.4]{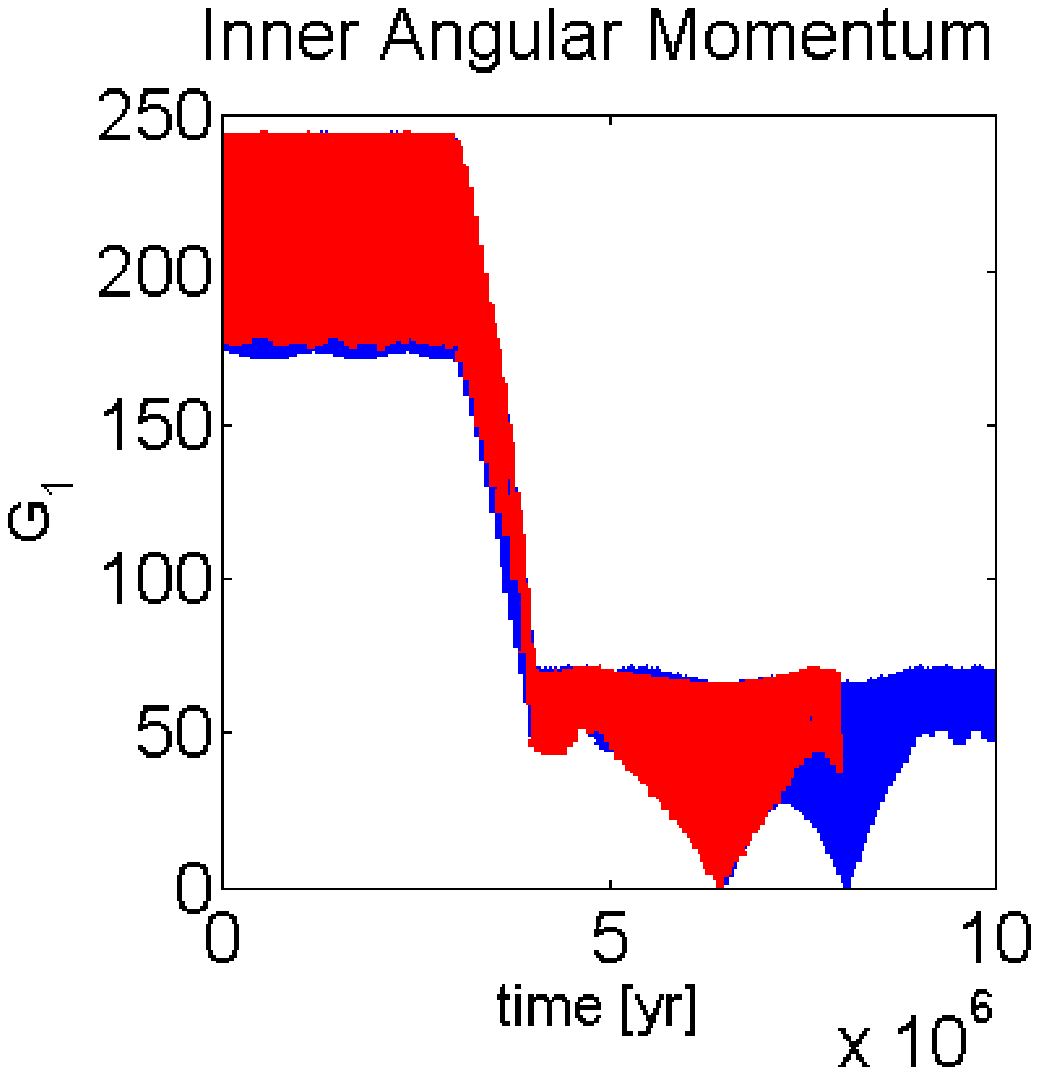}\includegraphics[scale=0.4]{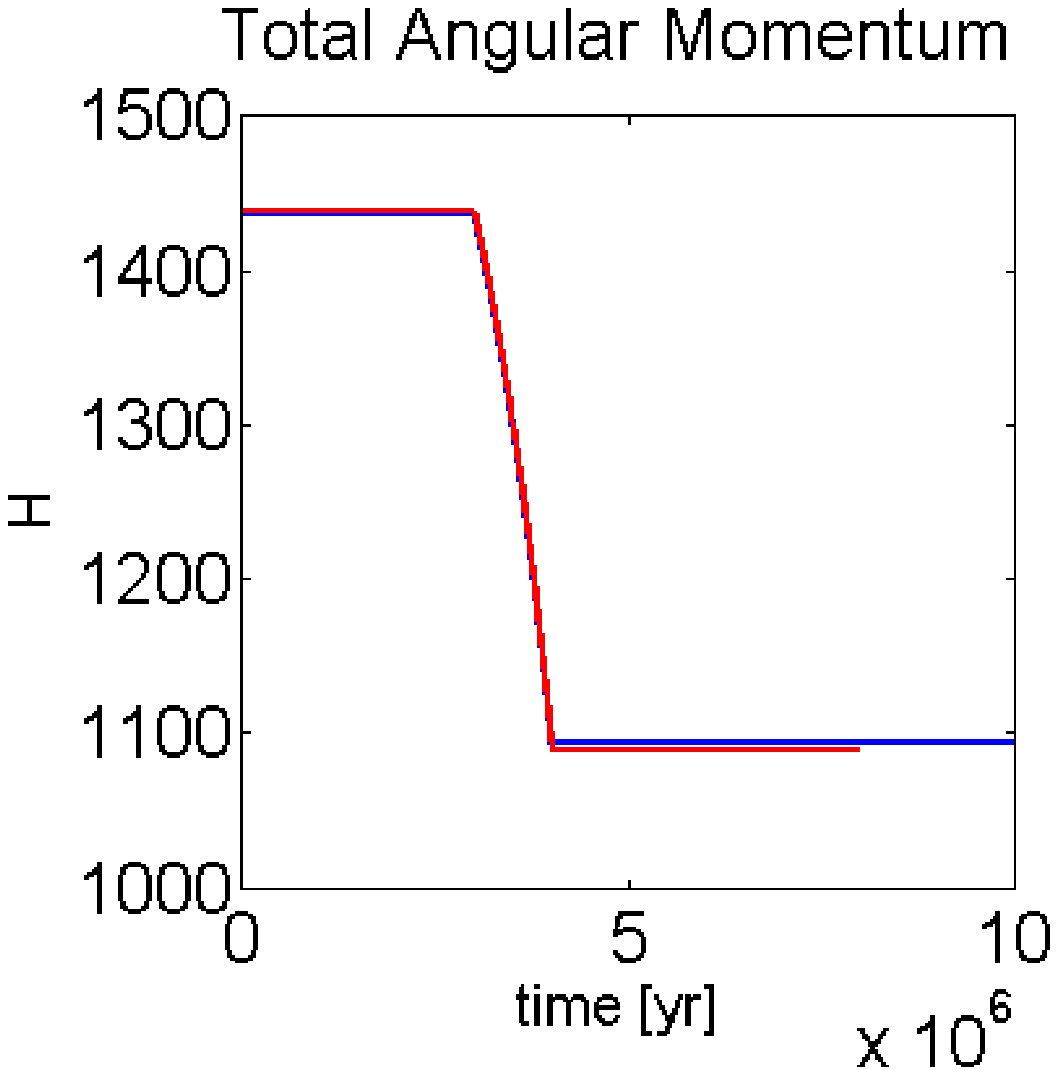}\caption{\label{fig:N_body}Similar evolution as shown in Fig. \ref{fig:MIEK},
comparing the secular method approach and N-body simulation. The secular
equation of motion is shown in blue, and the direct N-body integrations
is shown in red. Top left plot: inner eccentricity in $\log\left(1-e_{1}\right)$
scale. Top right: inclination vs. time. Bottom left: inner binary
angular momentum, $G_{1}$. Bottom right: total angular momentum vs.
time, $H\left(t\right).$ Excelent agreement can be seen between the
direct N-body simulation and the secular code.}
\end{figure*}

We integrated the system using both the secular method as well as
full N-body integration. The evolution of the systems is shown in
Fig. \ref{fig:MIEK}. We introduce a constant mass loss from $m_{0}$
starting after $t=3\,{\rm Myr}$ for a period of $\Delta t_{ml}=1$
Myr, until $m_{0}\left(t=4\,{\rm Myr}\right)=1.15M_{\odot}$. No mass
transfer was considered in this case, i.e. $\psi_{1,0}=\psi_{0,1}=\psi_{2,01}=0$,
and GR effects are negligible for this system. We can see very good
agreement between our secular method results and the direct N-body
simulations (which are themselves consistent with those obtained by
\citealp{Shappee2013}) as can be seen in Fig. \ref{fig:N_body}.

\begin{figure}[t]
\includegraphics[scale=0.5]{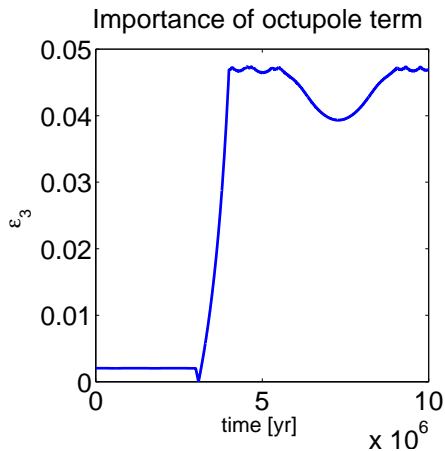}\caption{\label{fig:importance_octupole}Evolution of the $\epsilon_{3}$ parameter
(Eq. \ref{eq:importance of octupole}) as a function of time for the
same system as shown in Fig. \ref{eq:importance of octupole}. A dramatic
growth of the $\epsilon_{3}$ coefficient is noticeable after the
mass loss epoch at $t=4Myr$, by more than an order of magnitude.
The oscilatory behaviour of $\epsilon_{3}$ after mass loss is due
to small osilatory evolution of $e_{2}$ as one can eccpect from octupole
level of evulotion.}
\end{figure}

 As discussed in \citet{Shappee2013}, before mass loss, the system
is in the standard quadrupole Kozai mechanism regime, showing an oscillatory
behavior of the inclination and the inner eccentricity. At this stage
$\epsilon_{3}\approx0.002$, and the system is in the quadrupole regime,
while after mass loss $\epsilon_{3}$ becomes significantly larger
($\epsilon_{3}\approx0.045$; see Fig. \ref{fig:importance_octupole}).
This increase by more than an order of magnitude drives the system
to an octupole dominated regime, where the system can be driven into
much higher eccentricities; hence the term mass-loss induced eccentric
Kozai (MIEK).

\subsection{Mass loss/transfer from the third component and the secular evolution
freeze out (SEFO) process}

\label{sub:Mass-loss/transfer-from}
\begin{figure*}[!h]
\includegraphics[scale=0.4]{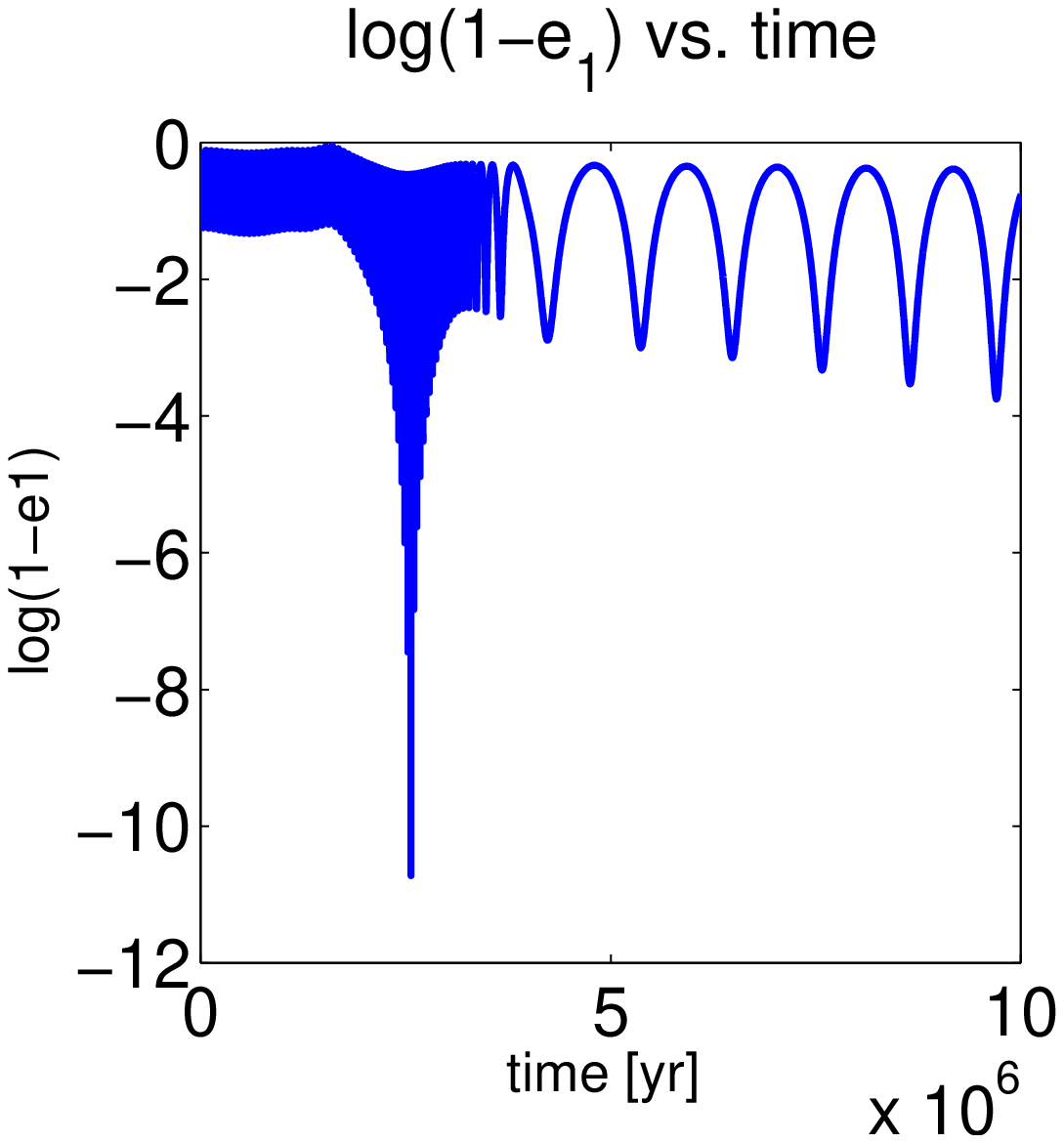}\includegraphics[scale=0.4]{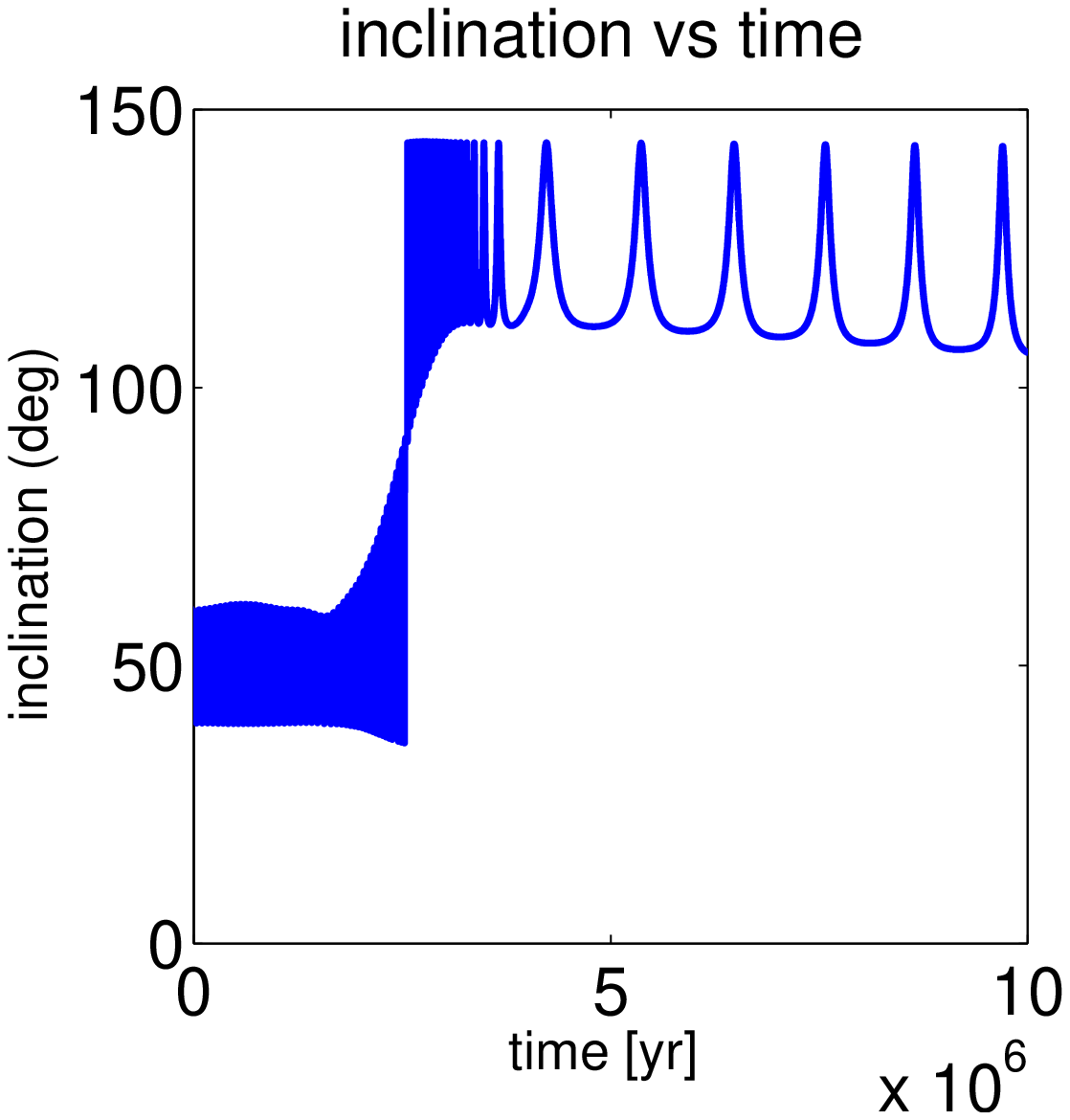}

\includegraphics[scale=0.4]{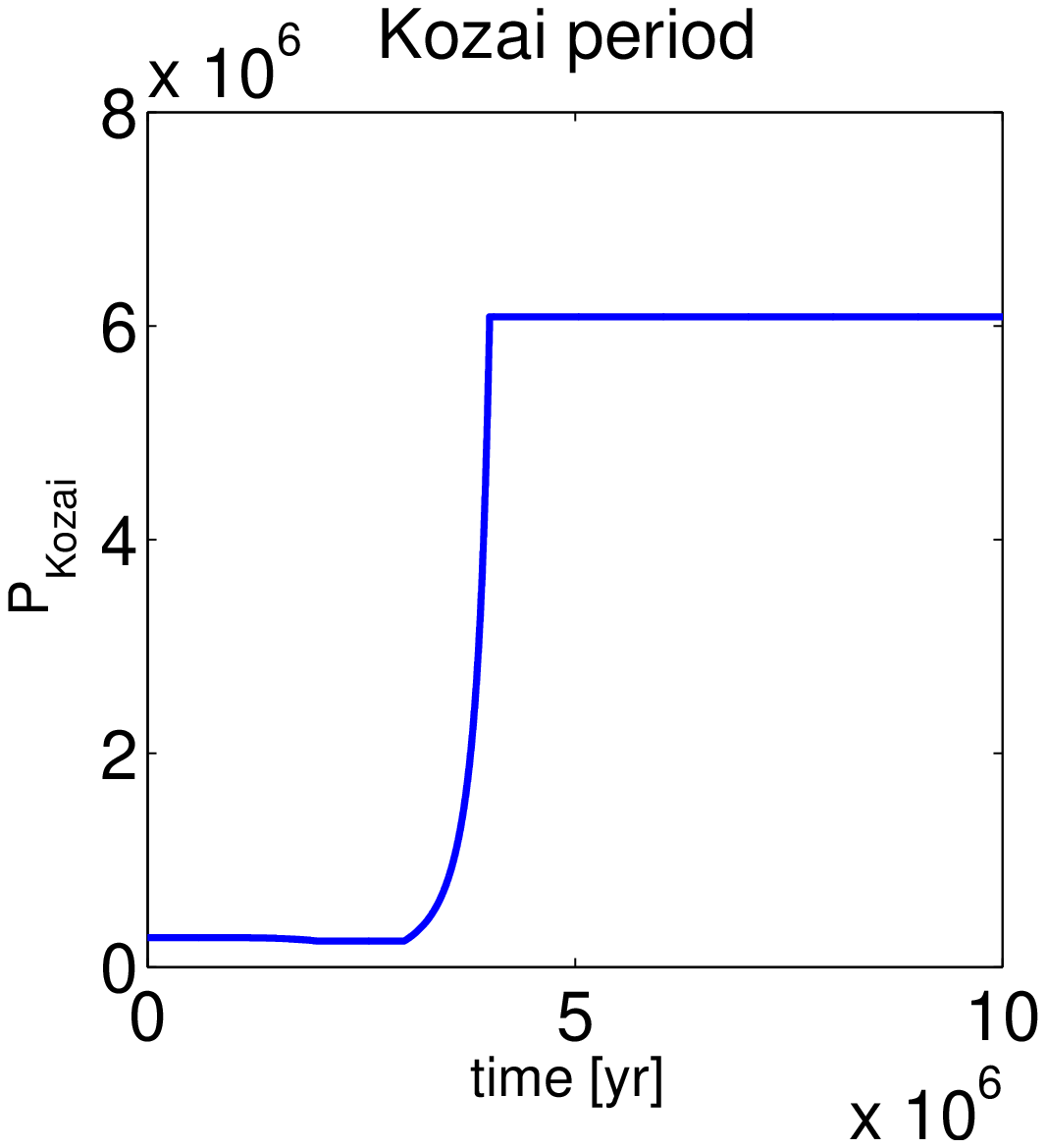}\includegraphics[scale=0.4]{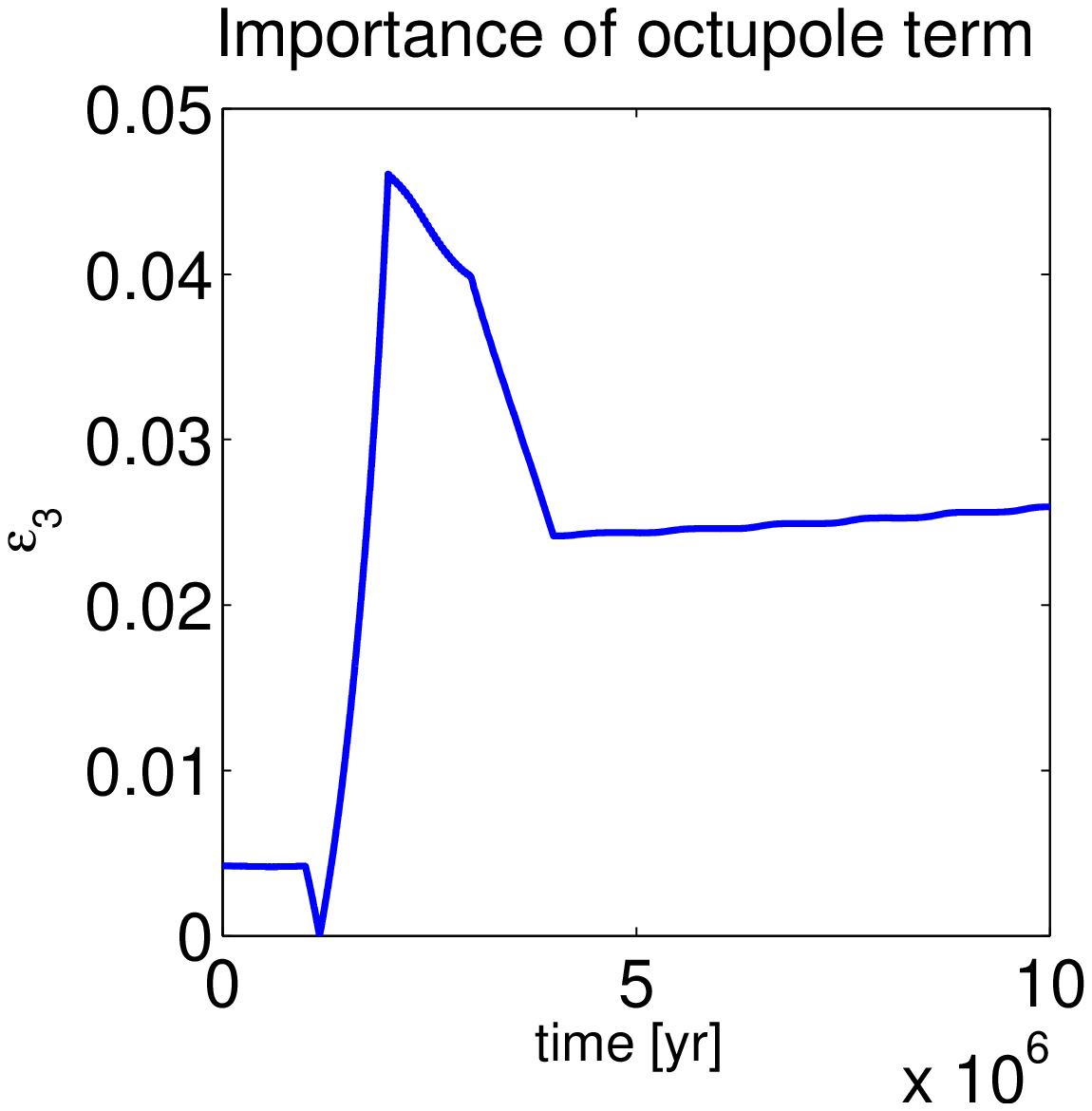}\caption{\label{fig:two_ML}Evolution of a triple system with mass loss from
both the inner binary primary component and a consecutive mass loss
from the thrid companion, showing a SEFO behaviour. The system paramters
are $m_{0}=7.0M_{\odot}$, $m_{1}=6M_{\odot}$, $m_{2}=6.5M_{\odot}$,
$a_{1}=10\left[AU\right]$, $a_{2}=250\left[AU\right]$, $e_{1}=0.1$,
$e_{2}=0.7$, $g_{1}=0^{\circ}$, $g_{2}=0^{\circ}$ and the mutual
inclination is $i=60^{\circ}$. The first constant mass loss from
$m_{0}$ is introduced after $t=1Myr$ for $\Delta t_{ml}=1Myr$ until
$m_{0}\left(t=2Myr\right)=1.15M_{\odot}$. A second constant mass
loss from $m_{2}$ is then introduced starting at $t=3Myr$ for $\Delta t_{ml}=1Myr$
until $m_{2}\left(t=4Myr\right)=1.15M_{\odot}$. Top right: mutual
inclination of the system. Top left: inner eccentricity as a function
of time in $\log\left(1-e_{1}\right)$ scale. Bottom left: $P_{Kozai}$
is plotted against time. Bottom right: $\epsilon_{3}$ is plotted
against time. \label{fig:Freeze_out_retrograde}The SEFO (secular
evolution freeze out) evolution can be seen in the transition occuring
furing the second mass-loss epoch, where $\epsilon_{3}$ becomes less
(by a factor of $\sim2$) and $P_{kozai}$ grows by an order of magnitude.}
\end{figure*}

We now consider the case of mass-loss
from the inner binary system of a triple (similar to the previous
case), but now also consider an additional later mass-loss epoch from
the third companion in the system. No mass-transfer is considered,
$\psi_{1,0}=\psi_{0,1}=\psi_{2,01}=0$ and GR effects are negligible.
Fig. \ref{fig:two_ML} shows the system evolution (see caption for
initial conditions).

The first mass loss epoch drives the system through a MIEK process
to the octupole regime, namely $\epsilon_{3}$ significantly grows
and the system evolves to be in the octupole level regime, where the
inner binary goes through a very eccentric orbit, and the inclination
evolves into a retrograde orbit. The second mass loss epoch drives
the system away from the octupole regime and closer into the quadrupole
regime. The system then seems to freeze in its current state, and
the inclination is kept on a retrograde orbit for a time longer than
the $t_{3}$ time scale (not shown in plots). This happens due the
mass loss, which changes both $P_{kozai}$ and $\epsilon_{3}$ in
such a way that the octupole level of evolution does not significantly
affect the system evolution on these long timescales, i.e. this process
leads to what we term a secular freeze-out (SEFO) 

\begin{figure*}[!h]
\includegraphics[scale=0.4]{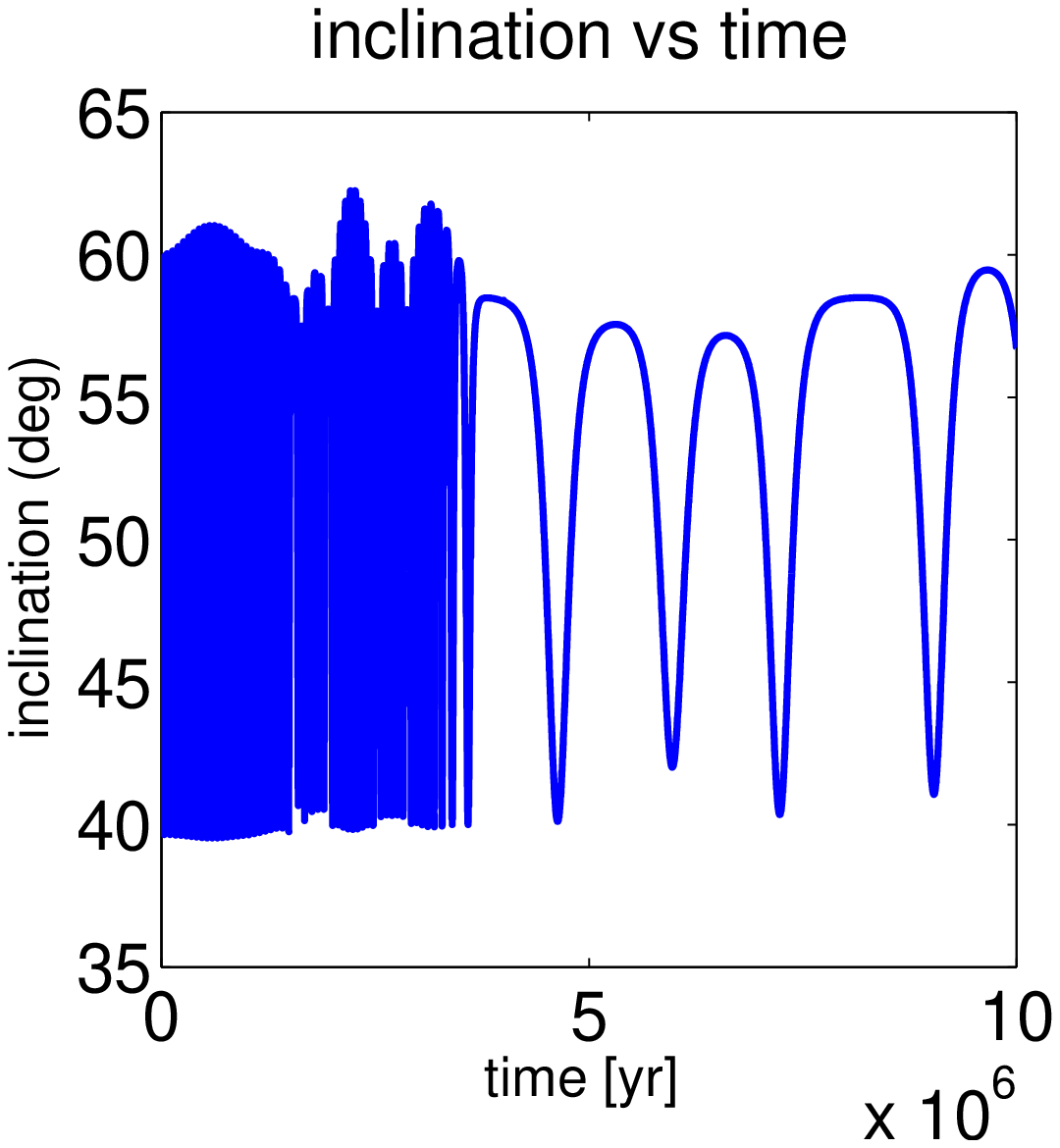}\includegraphics[scale=0.4]{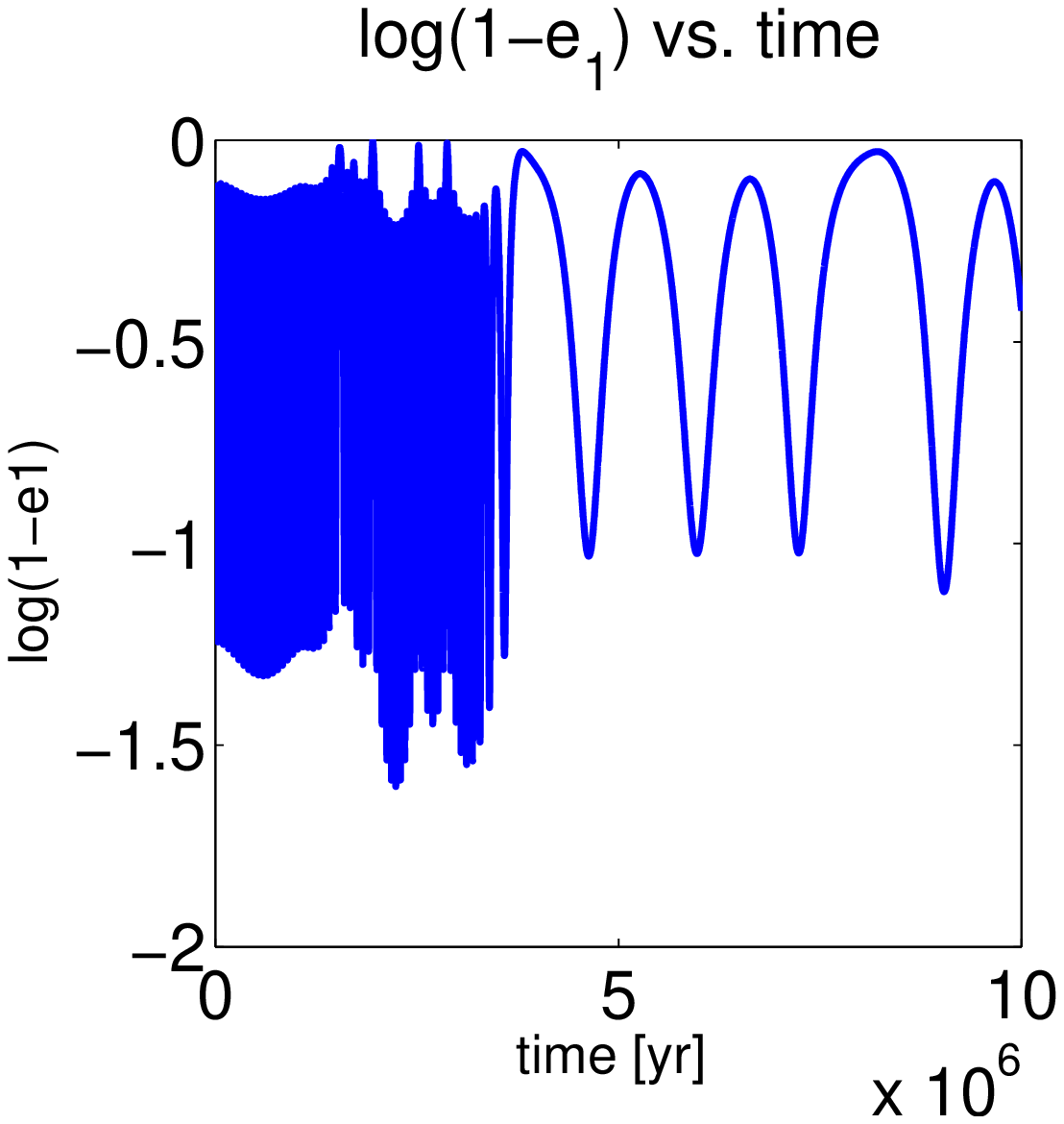}\caption{\label{fig:Retrograde_vs_NoRetrograde}Similar to Fig. (\ref{fig:Freeze_out_retrograde}),
but now showing an evolutionary case which includes mass transfer
in the inner binary, modeled with $\psi_{0,1}=0.5$. In this case
the system did not flip its inclination into a retrograde orbit during
its evolution in the octupole regime, and the later mass-loss from
the third companion led to the system secular freeze-out on a prograde
orbit. }
\end{figure*}

We now consider the effects of mass-transfer for the same system.
We integrated a triple with the same initial condition, but now we
allowed for mass transfer in the inner binary, namely $\psi_{0,1}=\kappa$,
for $\kappa<1$. For a wide range of $\kappa$ we got two distinct
outcomes: for $\kappa\lesssim0.45$ the end result was a freezed-out
retrograde system, while for $\kappa\gtrsim0.45$ the system did not
evolve into a retrograde orbit during its evolution. A representative
case for this mass-transfer evolution is shown in Fig. \ref{fig:Retrograde_vs_NoRetrograde}.

\subsection{Mass transfer from a third companion to the inner binary and the
formation of a short period inner binary}

\label{sub:Mass-transfer-from}
\begin{figure*}[!t]
\includegraphics[scale=0.4]{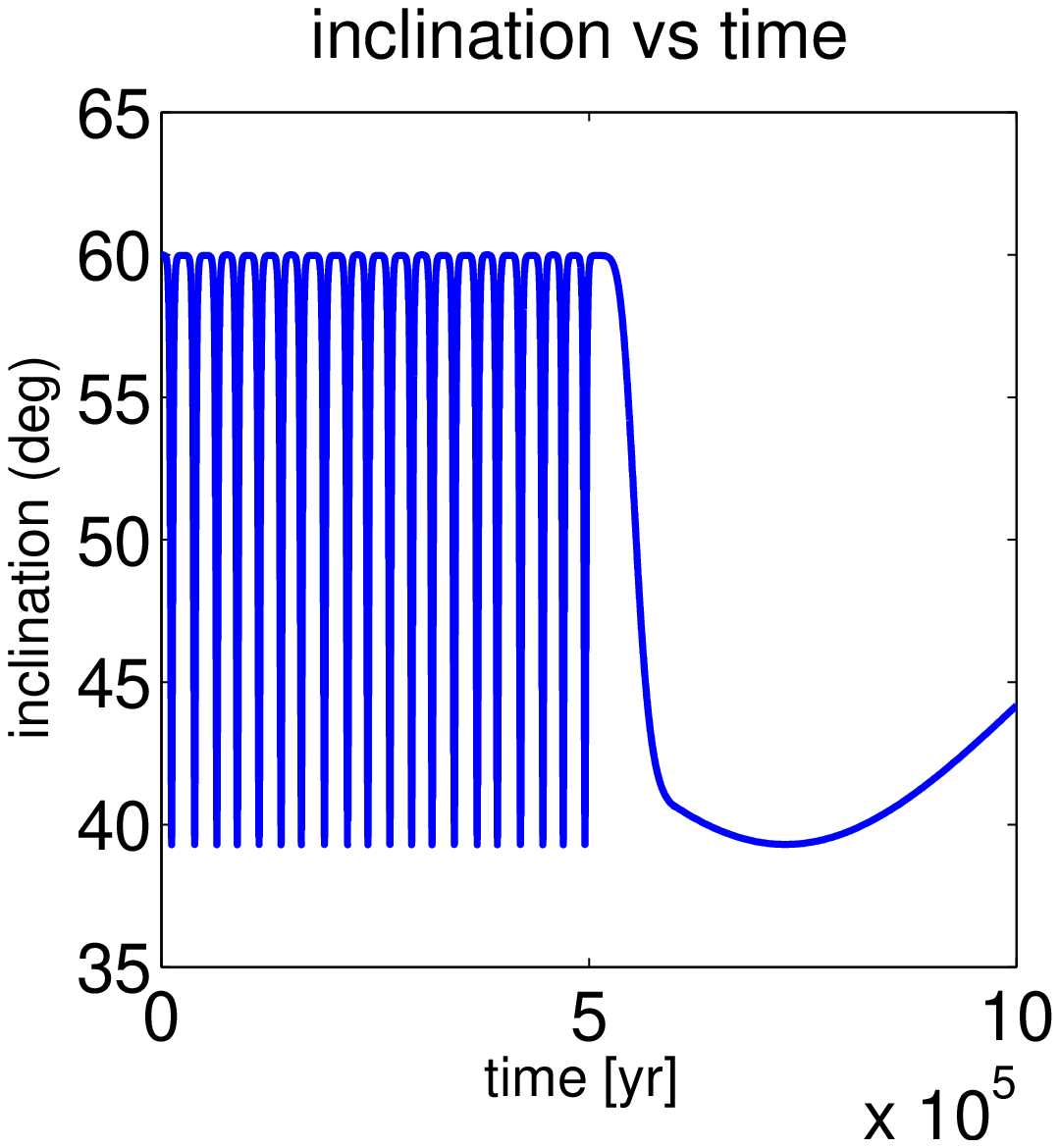}\includegraphics[scale=0.4]{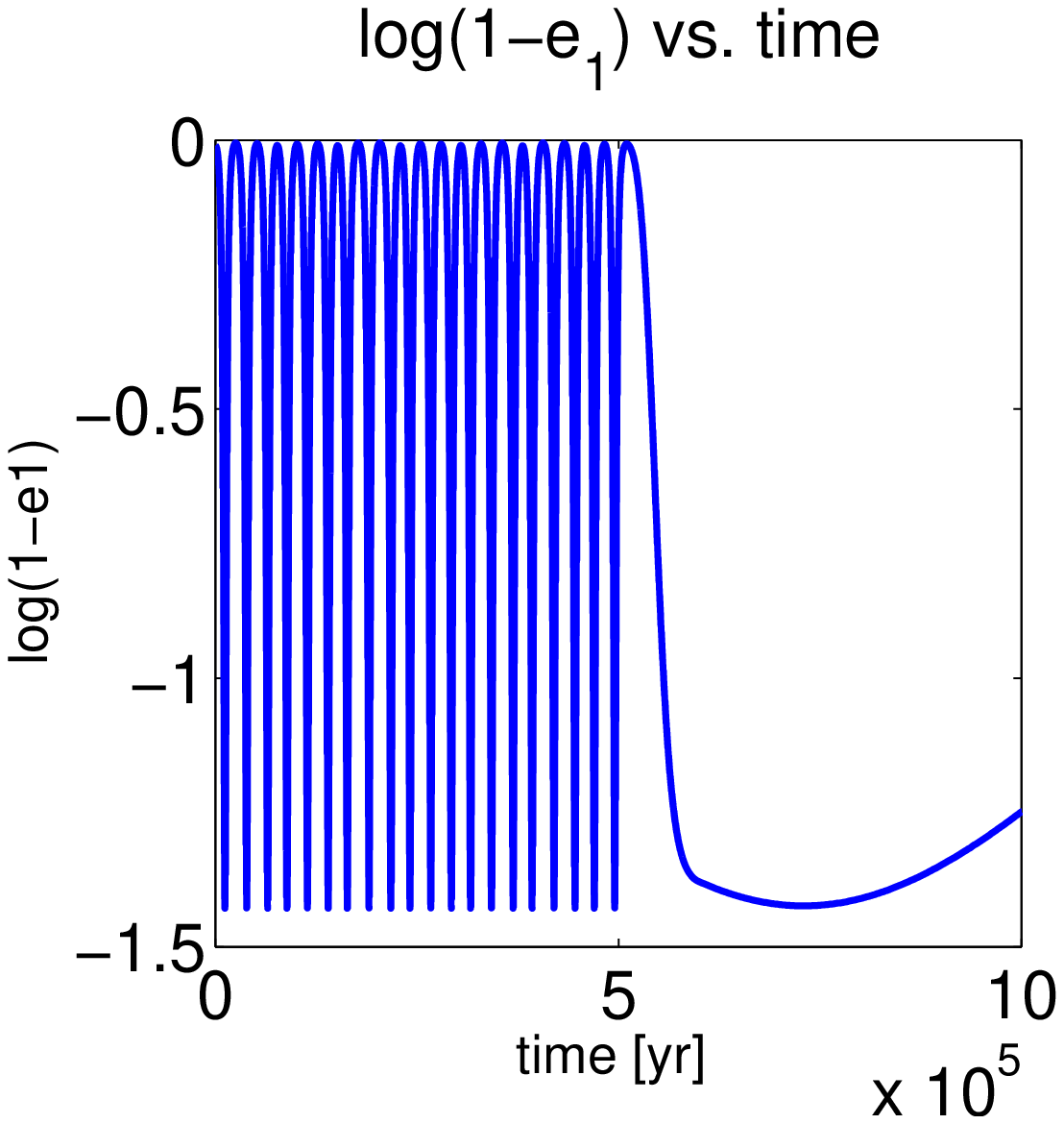}

\includegraphics[scale=0.4]{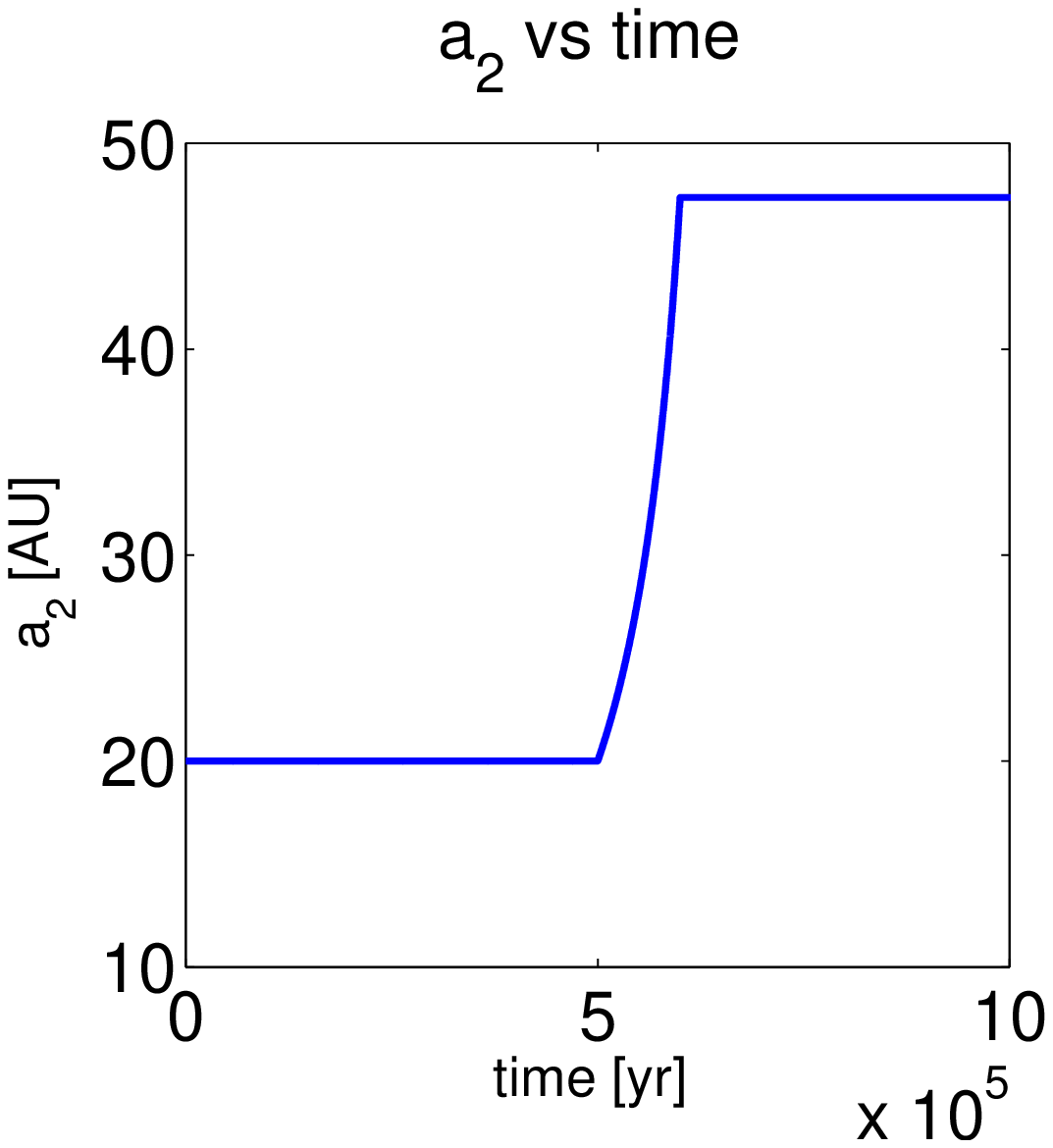}\includegraphics[scale=0.4]{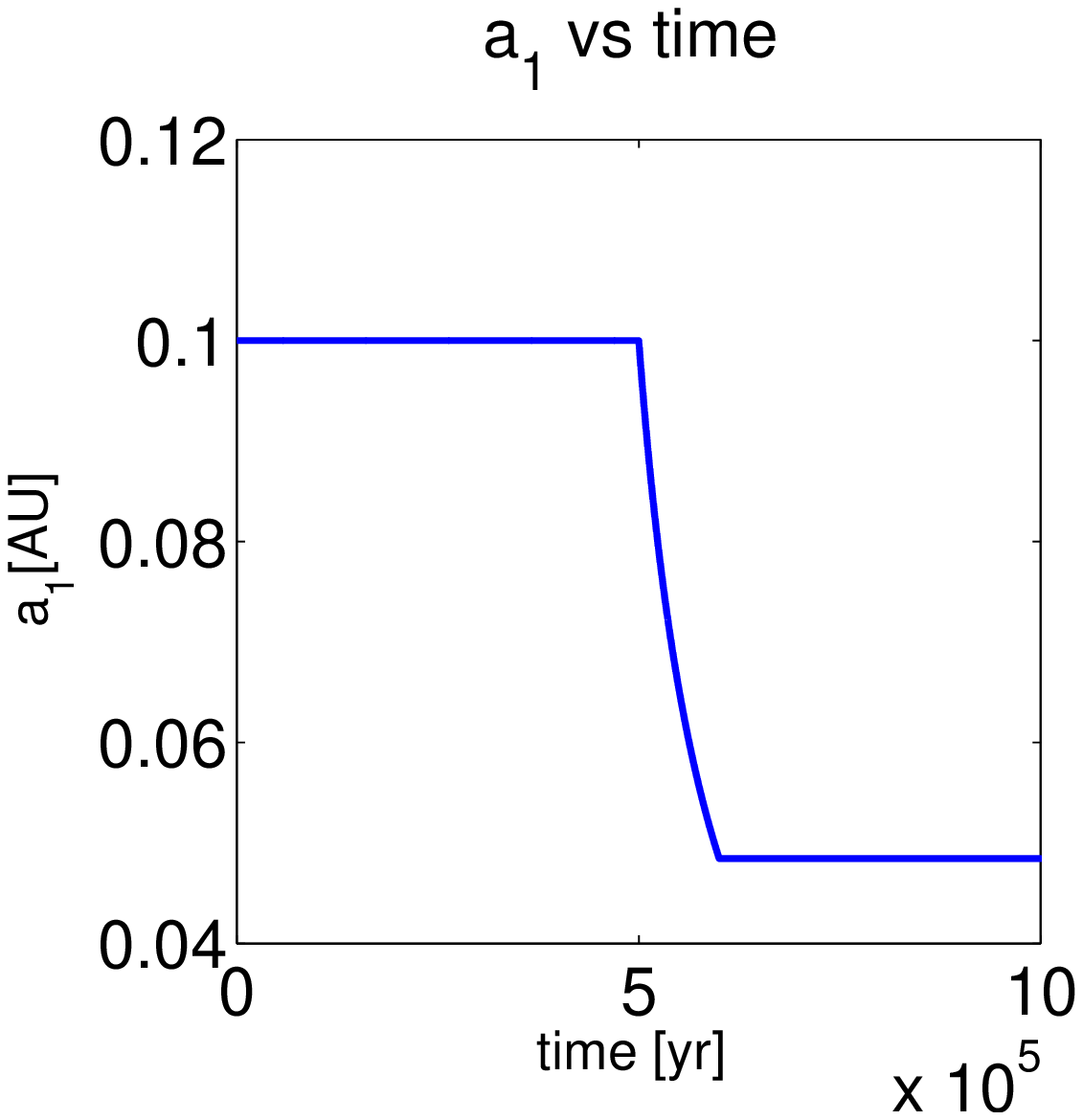}

\caption{\label{fig:MT_CB}The evolution of a triple system with mass loss
from the third companion to the inner binary. The system parameters
are $m_{0}=0.5\, M_{\odot}$, $m_{1}=0.6\, M_{\odot}$, $m_{2}=7\, M_{\odot}$,
$a_{1}=0.1\, AU$, $a_{2}=20\, AU$, $e_{1}=0.01$, $e_{2}=0.6$,
$g_{1}=0^{\circ}$, $g_{2}=0^{\circ}$ and $i=60^{\circ}$. A constant
secular mass loss from $m_{2}$ is introduced after $t=0.5\, Myr$
for $\Delta t_{ml}=10^{5}yr$ until $m_{2}=1.15\, M_{\odot}$, with
$\psi_{2,01}=0.2$. Top left: mutual inclination as a function of
time. Top right: inner binary eccentricity in $\log\left(1-e_{1}\right)$
scale. Bottom left: $a_{2}$ as a function of time. Bottom right:
$a_{1}$ as a function of time. After mass loss the inner binary shortens
to obtain an orbital period of $\sim3$ days.}
\end{figure*}

\begin{figure*}[!h]
\includegraphics[scale=0.4]{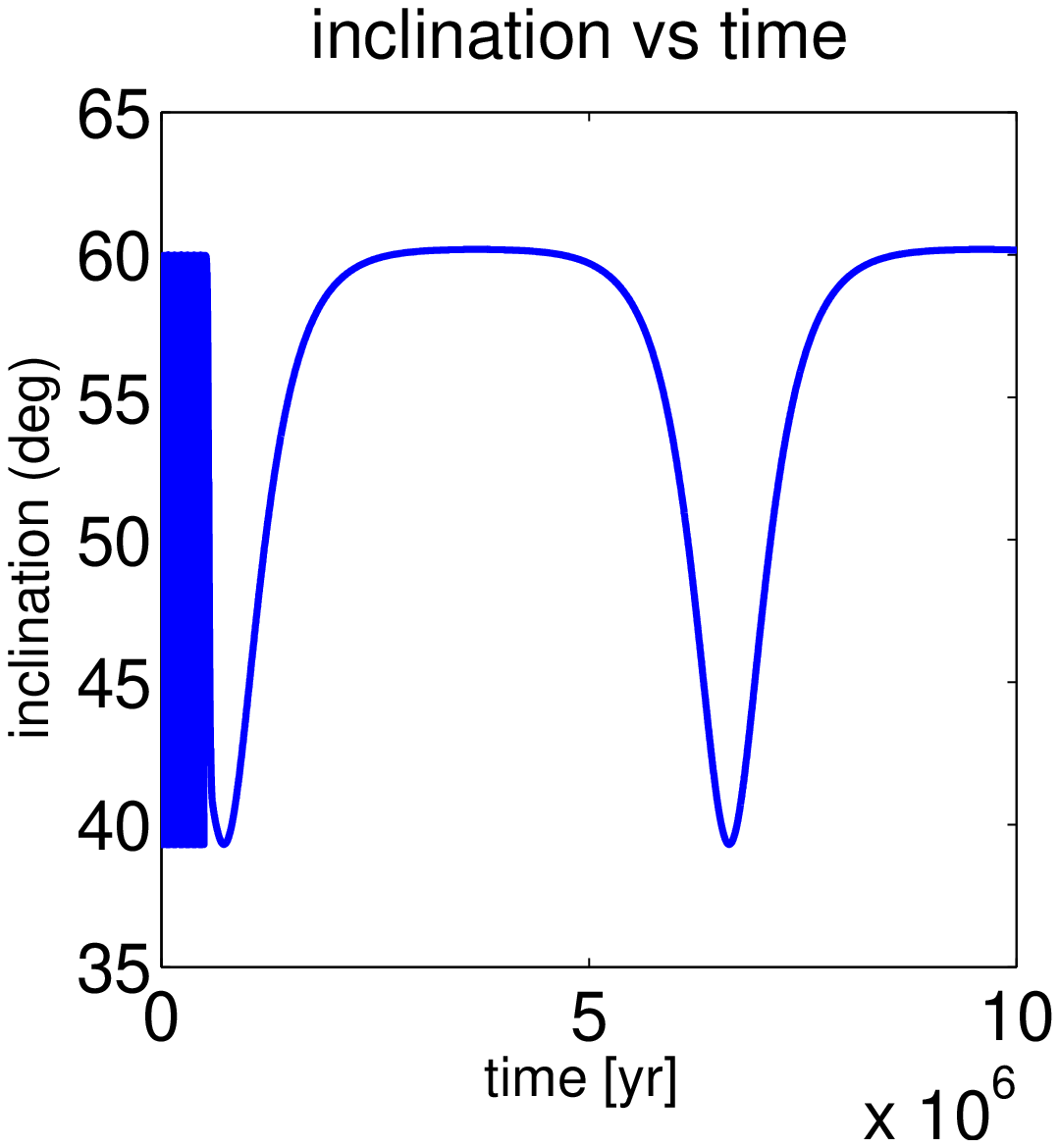}\includegraphics[scale=0.4]{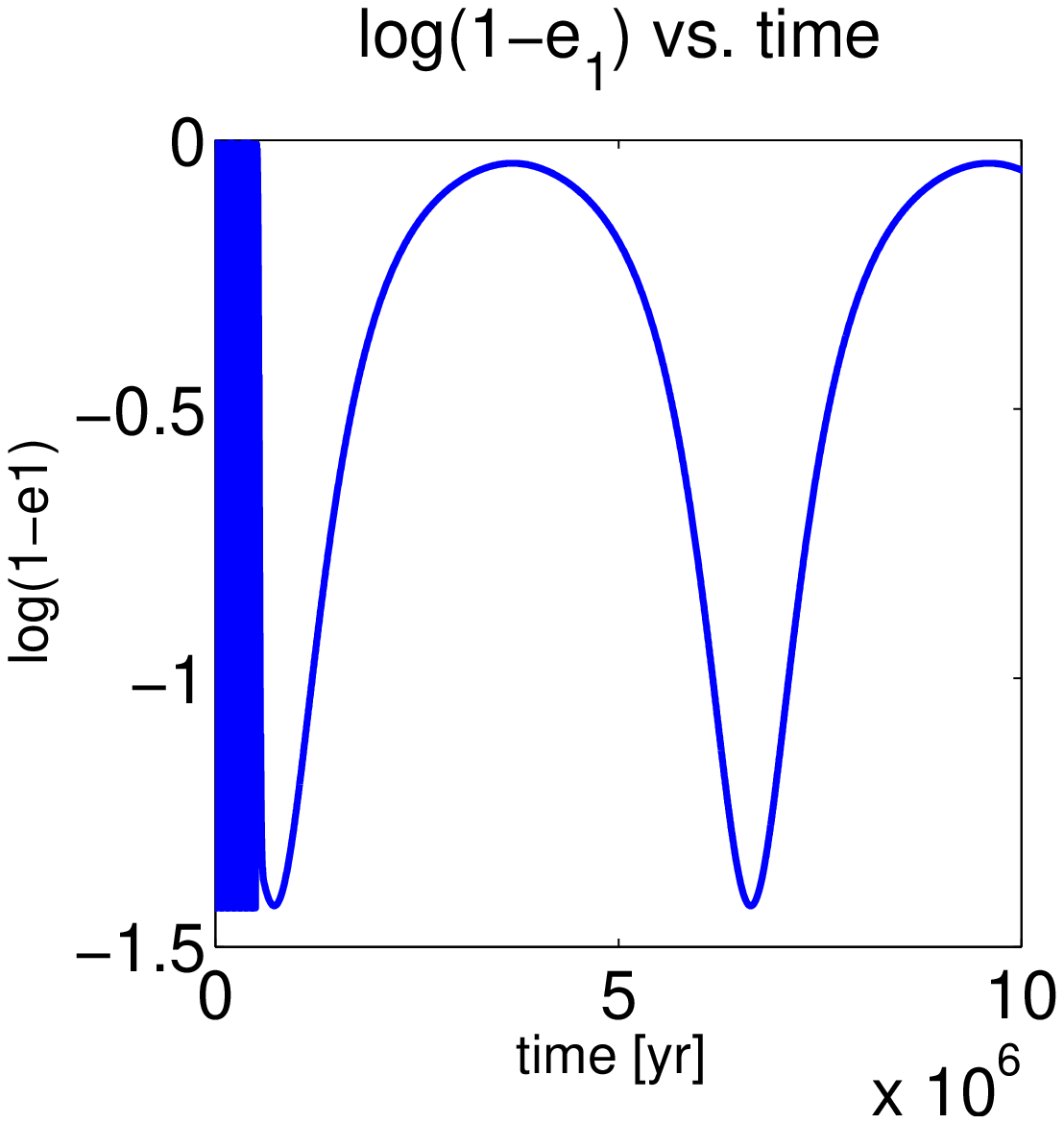}

\includegraphics[scale=0.4]{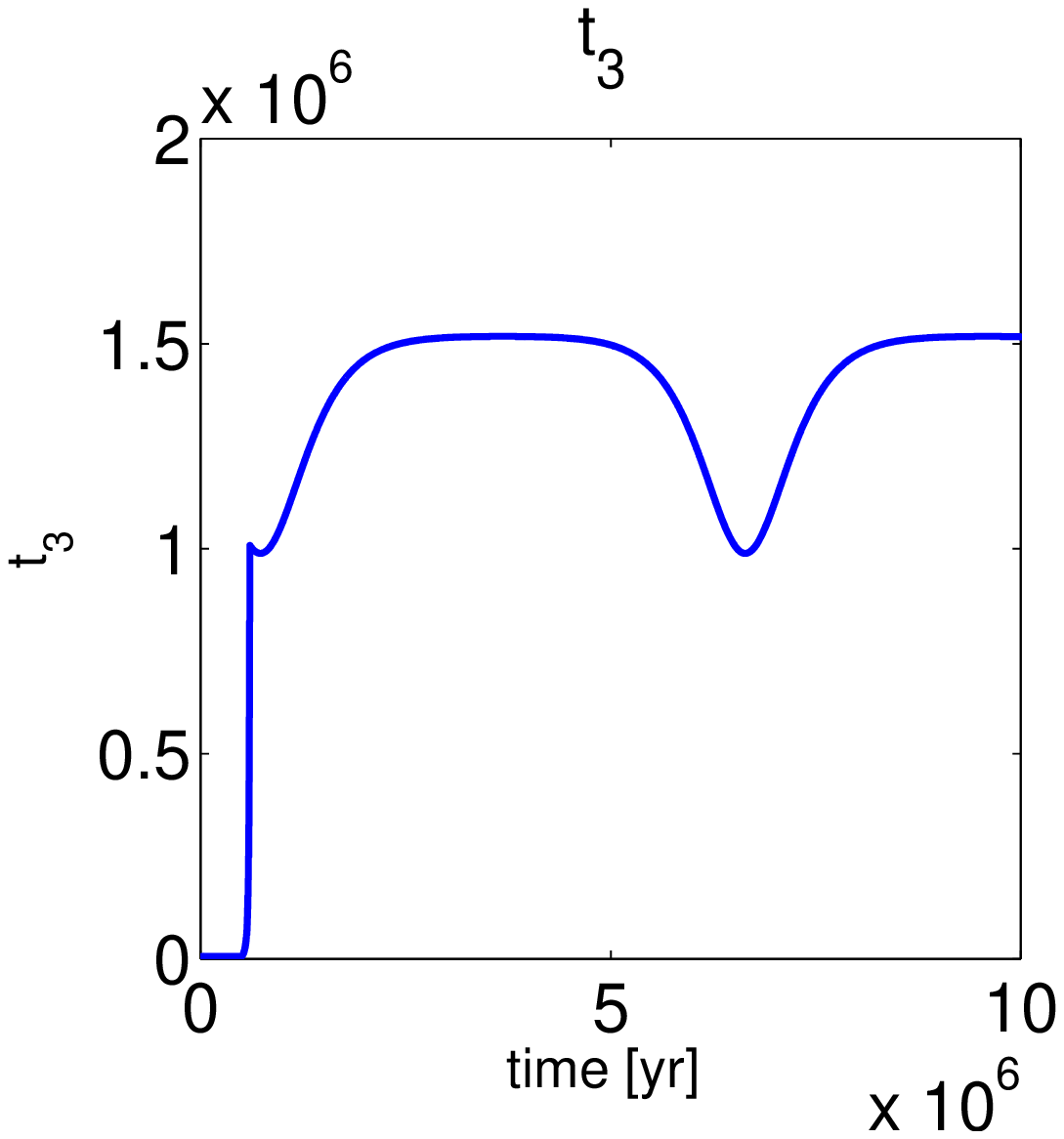}\includegraphics[scale=0.4]{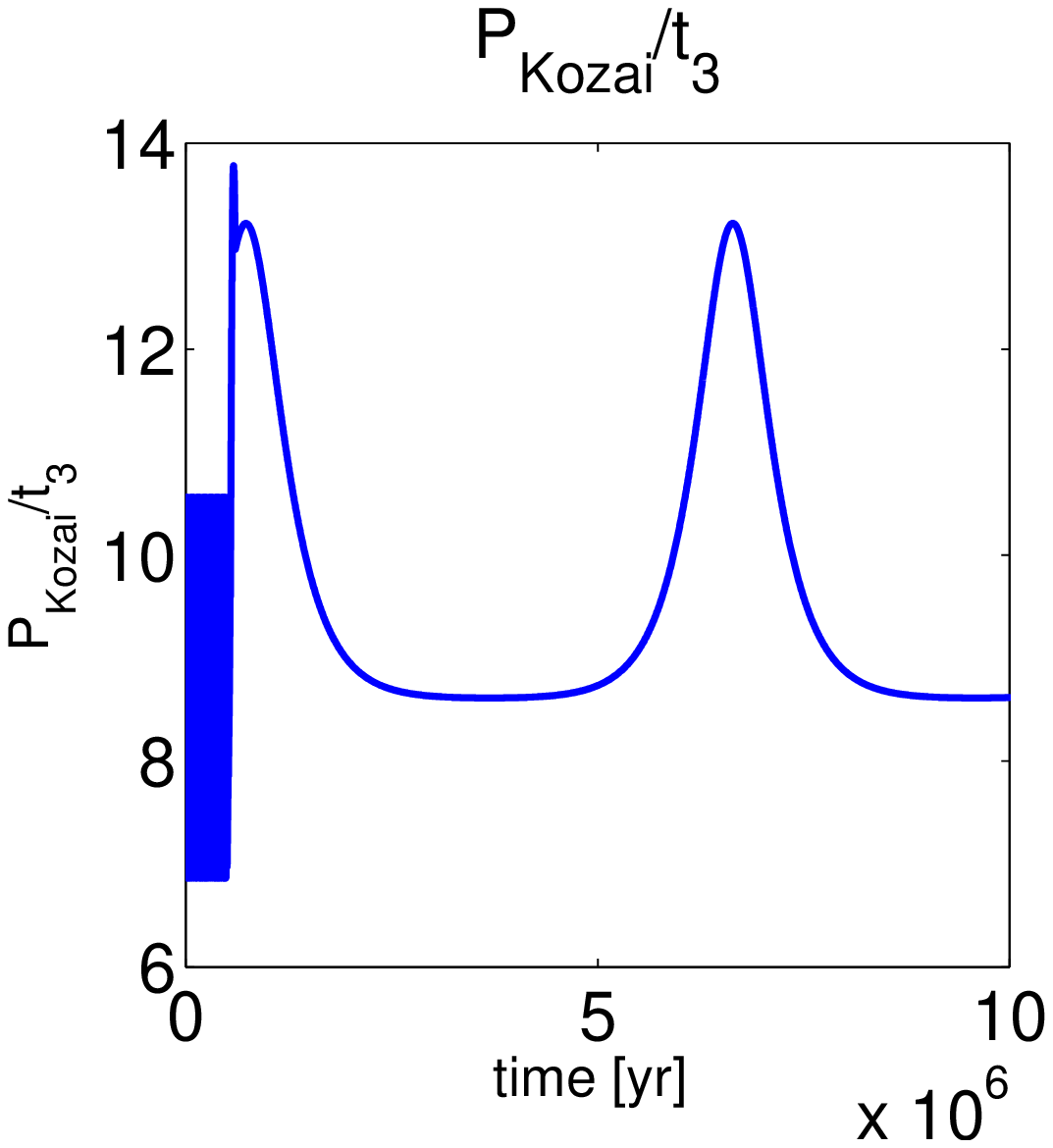}
\caption{\label{fig:MT_CB_LONG}Same as Fig. \ref{fig:MT_CB}, now showing
the long term evolution. Top left: mutual inclination as a function
of time. Top right: inner binary eccentricity in $\log\left(1-e_{1}\right)$
scale. Bottom left: octupole time scale, $t_{3}$ as a function of
time. Bottom right: ratio of Kozai period and the octupole timescale
as a function of time, $P_{Kozai}/t_{3}$.}
\end{figure*}

We now consider a case of mass-transfer
from the third companion into the inner binary. This mass-transfer
epoch increases the masses of the inner binary components and leads
to the compactification of the inner binary. We show an example of
such evolution in the following, where we also demonstrate the effects
of GR precession on the processes by showing the evolution of the
system both with and without including GR effects. In this case a
fraction of $\psi_{2,01}=0.2$ of the mass lost from $m_{2}$ is gained
by the inner binary; the mass is divided between the binary components
according to Eqs. (\ref{eq:m0_dot_mass_transfer}) and (\ref{eq:m1_dot_mass_trnasfer}).
Figs. \ref{fig:MT_CB} and \ref{fig:MT_CB_LONG} show the evolution
of the system on both short and long timescales, respectively, for
the case without GR effects (see system parameters in the figure caption).
As can be seen, after the mass transfer to the inner binary the inner
SMA shrinks by a factor of $\sim2$, and the period of the inner binary
reaches $\sim3$ days.

\begin{figure*}[!h]
\includegraphics[scale=0.32]{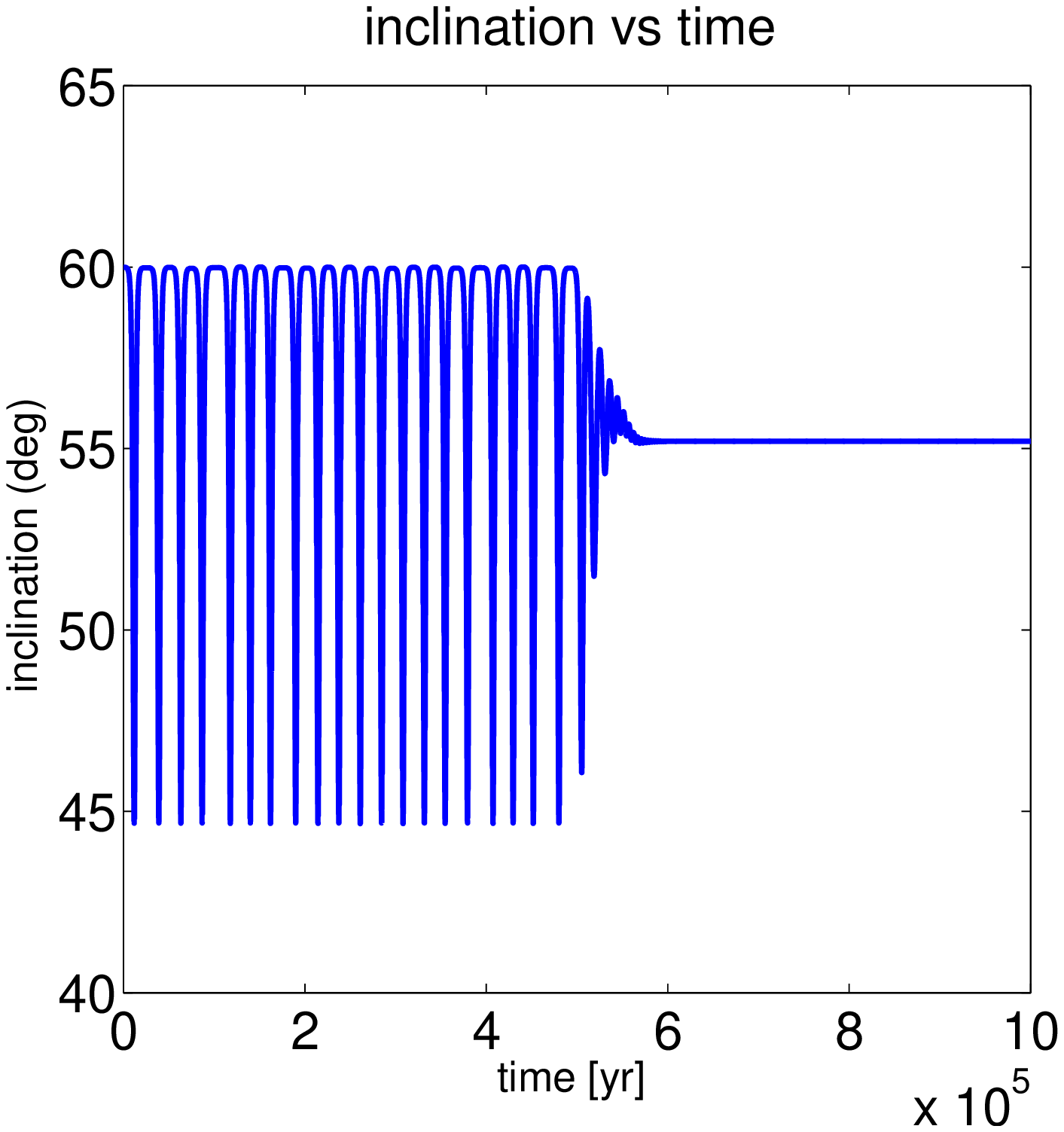}\includegraphics[scale=0.32]{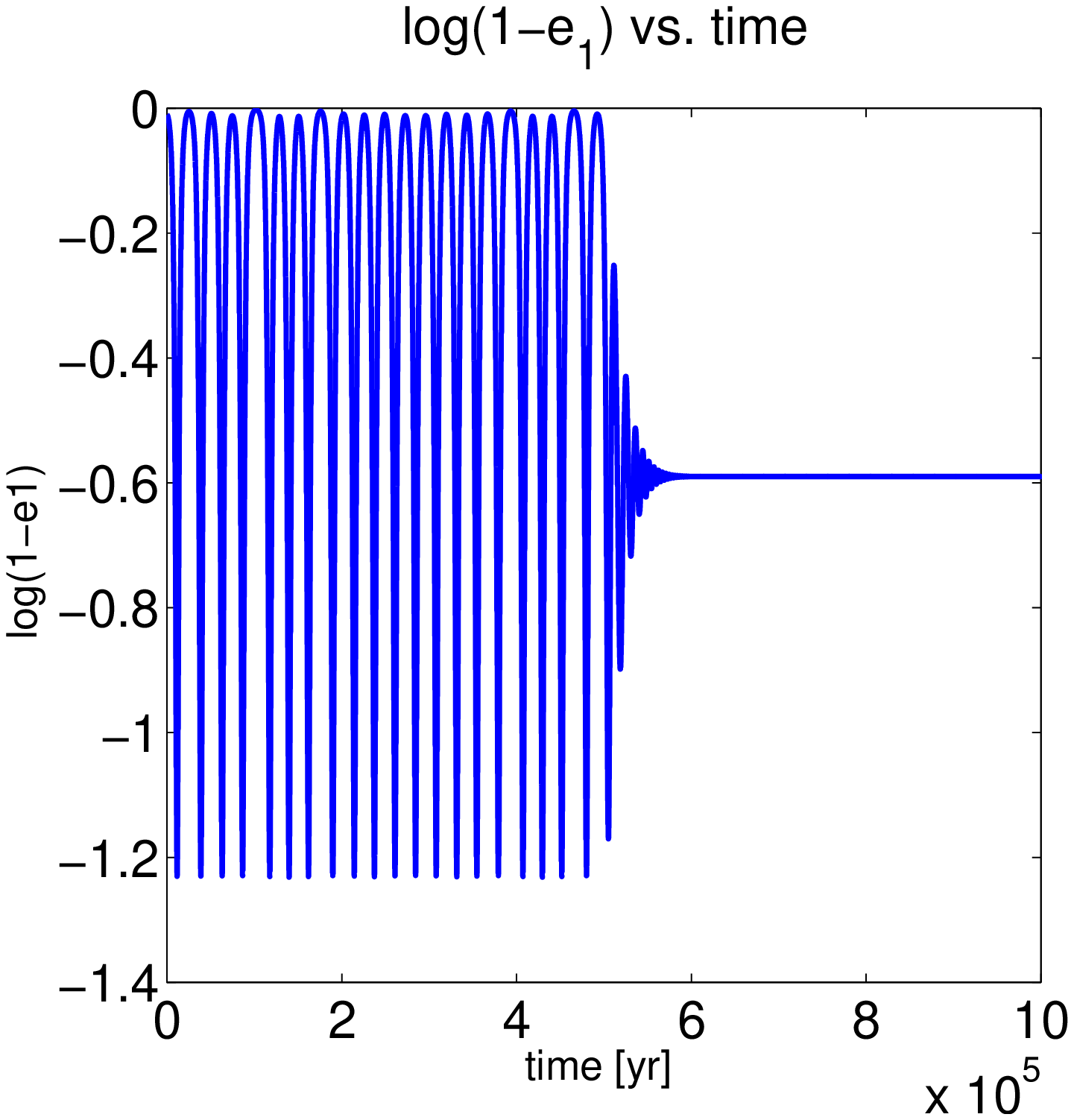}

\includegraphics[scale=0.32]{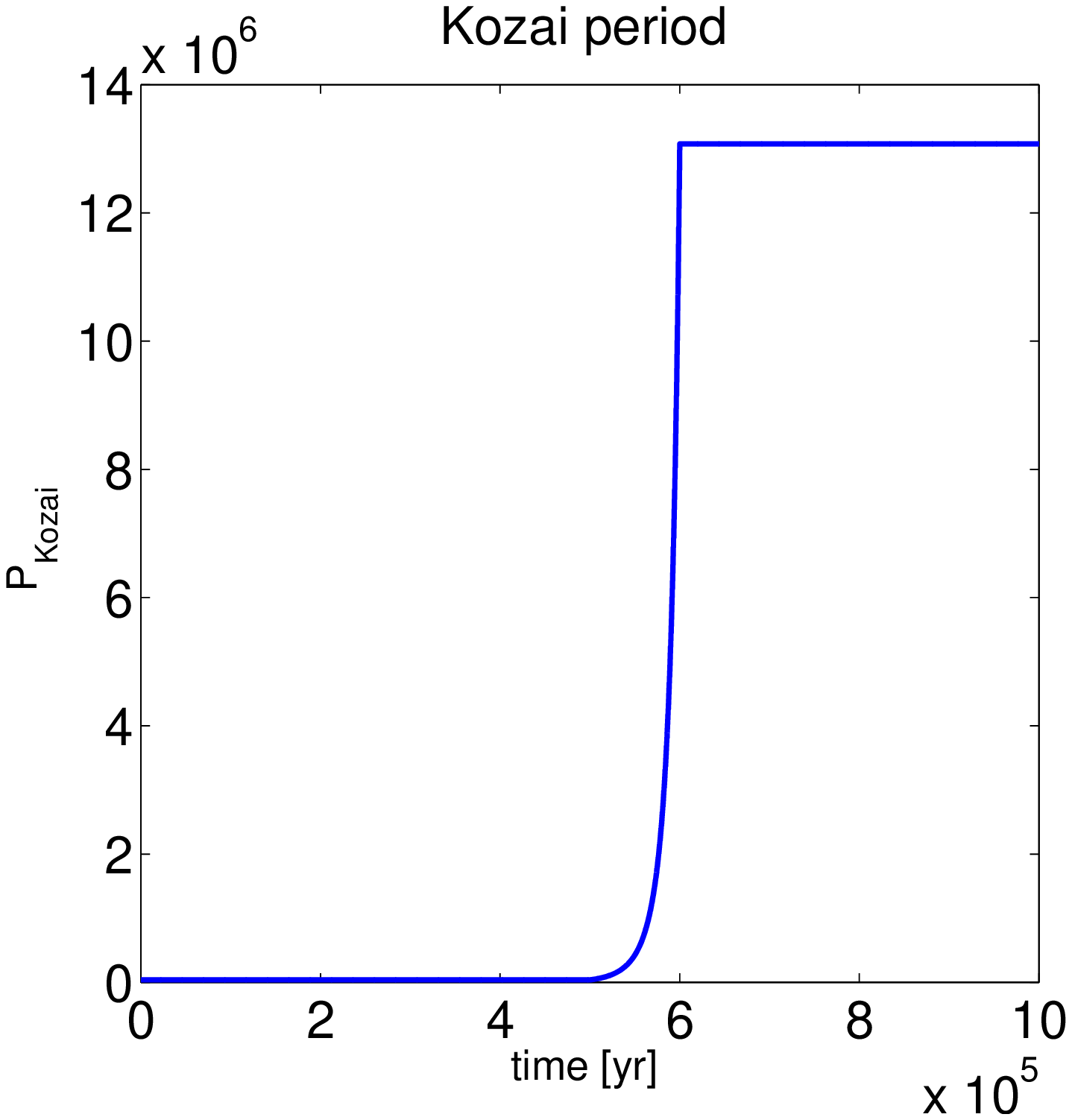}\includegraphics[scale=0.32]{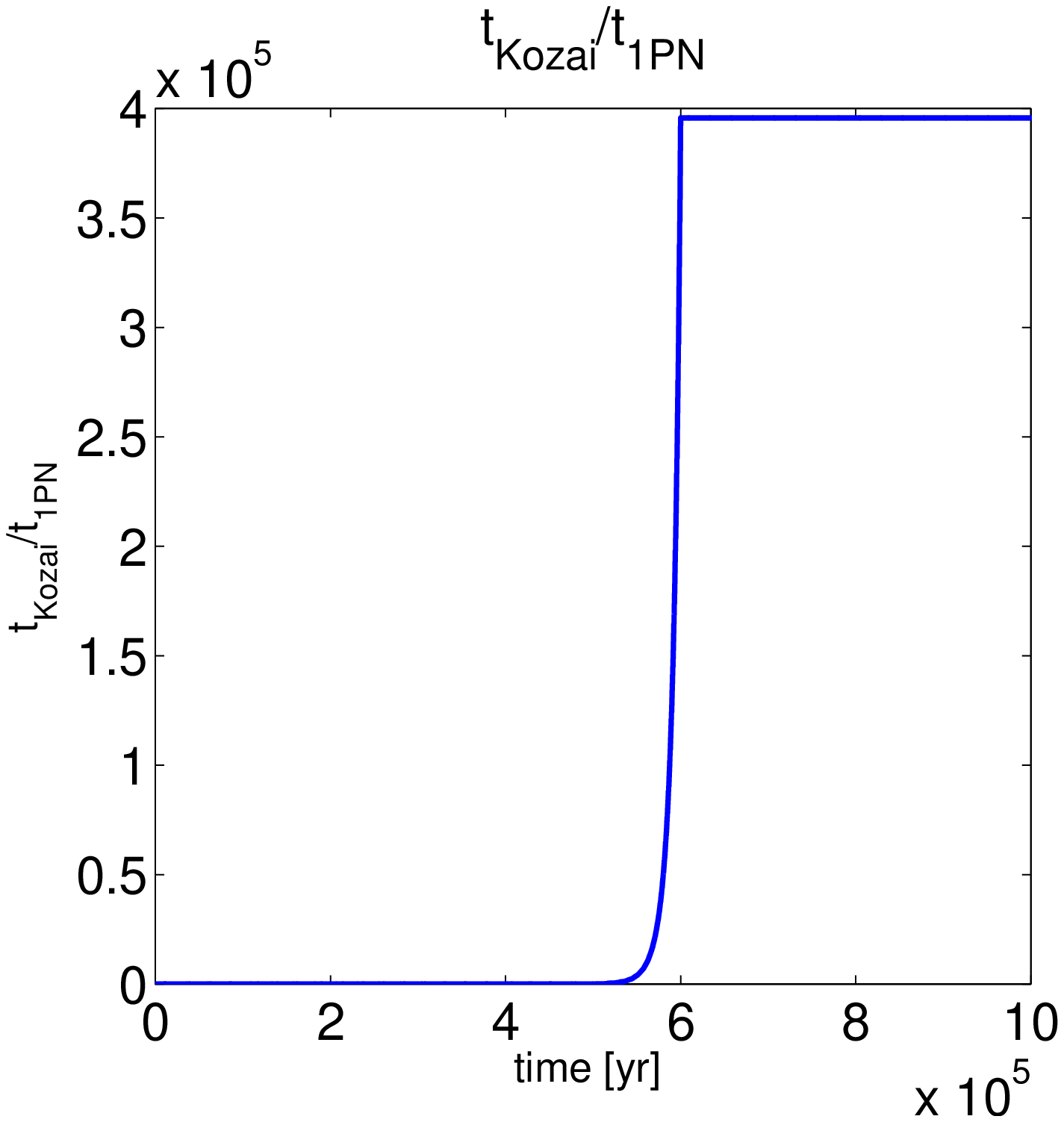}
\caption{\label{fig:MT_CB_GR}Similar to Fig. \ref{fig:MT_CB}, but including
GR effects. The system parameters: $m_{0}=0.5\, M_{\odot}$, $m_{1}=0.6\, M_{\odot}$,
$m_{2}=7\, M_{\odot}$, $a_{1}=0.1\, AU$, $a_{2}=20\left[\, AU\right]$,
$e_{1}=0.01$, $e_{2}=0.6$, $g_{1}=0^{\circ}$, $g_{2}=0^{\circ}$
and $i=60^{\circ}$. Constant secular mass loss on $m_{2}$ after
$t=0.5\, Myr$ for $\Delta t_{ml}=10^{5}yr$ until $m_{2}=1.15\, M_{\odot}$,
$\psi_{2,01}=0.2$, GR effects are included. Top left: mutual inclination
as a function of time. Top right: inner binary eccentricity in $\log\left(1-e_{1}\right)$
scale. Bottom left: $P_{Kozai}$ as a function of time. Bottom right:
$P_{Kozai}/t_{1PN}$ as a function of time. After mass transfer $t=0.6\, Myr$
the percession timescale is much shorter than the Kozai period resulting
in quenching the Kozai effect.}
\end{figure*}

Fig. \ref{fig:MT_CB_LONG} shows the evolution of the mutual inclination
and the inner binary eccentricity on a longer timescale compared with
Fig. \ref{fig:MT_CB}. On this longer timescale the modulation from
the octupole perturbation regime can also be seen.

GR effects become important when the timescales of $t_{1PN}$ and
$t_{2.5PN}$ become comparable or smaller than the other dynamical
timescales. This happens when the inner binary becomes highly eccentric
and/or the inner binary SMA, $a_{1}$ becomes small. Once the timescale
for $t_{1PN}$ becomes comparable to the $P_{Kozai}$, GR precession
perturbs the coherent evolution of the Kozai-Lidov evolution and the
amplitude of the oscillatory behavior of the Kozai mechanism\citep{Naoz2013a}
is quenched. In Fig. \ref{fig:MT_CB_GR} we show the evolution of
the same system, where we now include GR effects; the quenching of
the oscillatory behavior by the GR effects is clearly seen and can
be well understood, since after the mass-transfer the 1PN timescale
becomes smaller than the Kozai period. Indeed, the Kozai period increases
by more than an order of magnitude up to $10\, Myr$, and we find
no variation of the orbital elements occurring on this (or longer)
timescale (not shown). 

\section{Discussion and Summary}

\label{sec:Discussion-and-Summary} Realistic stellar systems undergo
different evolutionary scenarios where secular mass loss and mass-transfer
play an important role. These include winds from evolved stars, Roche
lobe overflow (RLOF), wind RLOF etc. Incidents of mass-transfer or
mass-loss may also be relevant to other systems, whether planets,
moons or asteroidal systems (e.g. through atmospheric evaporation).
Such processes have been studied extensively and their effects on
binary systems have been explored. Here we extended the study of these
effects and model the effects of secular mass-loss/transfer on the
long term dynamical behavior of \emph{triple} systems; the rich dynamics
of triples becomes even more complex when coupled to such variation
in the mass of the components.

Here we have shown for the first time that the dynamics of a triple
hierarchical system including mass-loss and mass-transfer can be well
modeled via double averaged Hamiltonian expanded up to an octupole
level \citep{Harrington1968,Ford2000,Blaes2002}. This Hamiltonian
presents several relevant timescales, which are significantly dependent
on the mass and separation of the system components. Mass-loss/mass-transfer
therefore affect these relevant time-scales and can thereby transfer
a triple system from one type of dynamical behavior regime to another,
where the dynamics are dominated by a different type of perturbations/physical-processes. 

We used our model to study the evolution of various types of triple
systems, and demonstrated several such evolutionary channels. These
include the mass-loss induced eccentric Kozai (MIEK) process (also
studied by \citet{Shappee2013} using N-body integration) and its
reversed inverse-MIEK process, the secular evolution freeze-out (SEFO)
and mass transfer from a third companion to an inner binary (see also
related study by SPH and N-body integration by \citet{deVries2014}).
The MIEK process transfers a triple from the quadrupole (stable oscillatory
behavior) to the octupole regime (quasi-periodic and even chaotic
behavior leading to extremely large changes in eccentricities and
inclinations), where an inverse-MIEK, transferring a system from the
octupole to the quadrupole regime, can also occur when mass is lost
from the third outer companion. An additional similar process, SEFO,
can transfer a system from the quadrupole regimes to a state where
secular evolution is either quenched or operates on excessively long
time-scales. Mass transfer from a third companion can both induce
the formation of short-period binaries, as well as lead to quenched
secular evolution, at the point where the systems becomes more susceptible
to GR effects, which quench any further oscillatory behavior. 

Finally, the model developed here shows excellent agreement with full
N-body integration schemes, but has the important advantage of providing
much faster (orders of magnitude) calculations. It provides a useful
tool for the study of triple systems, and in particular for the exploration
of a large phase space of initial conditions. It could therefore also
be integrated into studies of larger-scale systems (such as stellar
clusters, where triples can play an important role) and/or population
synthesis studies where understanding the evolution of many triple
systems becomes important.

\acknowledgements{We acknowledge support from the I-CORE Program of the Planning and
Budgeting Committee and The Israel Science Foundation grant 1829/12.
HBP is a Deloro and BIKURA fellow. }
\newpage
\bibliographystyle{apj}

\end{document}